\documentclass[aps,preprintnumbers,showpacs,superscriptaddress,groupedaddress]{revtex4}  

\usepackage{amssymb}
\usepackage{amsmath}
\usepackage{amssymb}
\usepackage{graphicx}
\usepackage{subfigure}
\usepackage{color}
\usepackage{mathrsfs}
\usepackage[dvipsnames]{xcolor}

\usepackage[breaklinks=true,colorlinks=true]{hyperref}
\hypersetup{colorlinks=true,citecolor=blue,linkcolor=blue,urlcolor=blue}

\usepackage[utf8]{inputenc}
\usepackage[english]{babel}

\begin{document}

\preprint{\small Eur.\ Phys.\ J.\ C \textbf{81}, 1124 (2021)   
\ [\href{https://doi.org/10.1140/epjc/s10052-021-09935-7}{DOI: 10.1140/epjc/s10052-021-09935-7}]
}

\title{Exotic final states in the $\varphi^8$ multi-kink collisions}

\author{Vakhid A. Gani}
\email{vagani@mephi.ru}
\affiliation{Department of Mathematics, National Research Nuclear University MEPhI\\ (Moscow Engineering Physics Institute), Moscow 115409, Russia}
\affiliation{Theoretical Department, National Research Center Kurchatov Institute, Institute for Theoretical and Experimental Physics, 117218 Moscow, Russia}
\author{Aliakbar Moradi Marjaneh}
\email{moradimarjaneh@gmail.com}
\affiliation{Department of Physics, Quchan Branch, Islamic Azad university, Quchan, Iran}

\author{Kurosh Javidan}
\email{javidan@um.ac.ir}
\affiliation{
Department of Physics,  Faculty of Sciences, Ferdowsi University of Mashhad, Mashhad, Iran}

\begin{abstract}

We study final states in the scattering of kinks and antikinks of the $\varphi^8$ field-theoretic model. We use the initial conditions in the form of two, three or four static or moving kinks. In the numerical experiments we observe a number of different processes such as emergence of static and moving oscillons, change of the kink's topological sector, scattering of an oscillon by a kink, production of kink-antikink pairs in oscillon-oscillon collisions. In antikink-kink collisions for asymmetric kinks, we found resonance phenomena --- escape windows.

\end{abstract}



\maketitle

\section{Introduction}
\label{sec:Introduction}

Many physical systems are described in terms of field-theoretic models with a real scalar field, the dynamics of which is determined by nonlinear partial differential equations \cite{Vilenkin.book.2000,Manton.book.2004,Vachaspati.book.2006}. Among the field configurations that satisfy the equations of motion, topologically nontrivial objects, including the so-called {\it kinks} (see, e.g., \cite[Ch.~5]{Manton.book.2004} and \cite{Vachaspati.book.2006}), are of particular importance. The study of the properties of kink solutions is of great interest for various physical applications.

This paper is devoted to the study of new phenomena in the scattering of kinks of the $\varphi^8$ model. The history of the numerical study of interactions of kink solutions of nonlinear partial differential equations goes back more than half a century. Nevertheless, in recent years, the number of new results in this area has not decreased; see, e.g., \cite{Kevrekidis.book.2014,Shnir.book.2018,Kevrekidis.book.2019} for review.

The $\varphi^8$ model is one of the widely used models with polynomial self-interaction (potential) of a scalar field. In connection with applications, several different modifications of this model are considered \cite{Lohe.PRD.1979,Khare.PRE.2014,Gani.JHEP.2015,Christov.PRD.2019,Gani.PRD.2020}. Depending on a specific form of the potential, the model exhibits different properties. In particular, the model can have different sets of topological sectors, different asymptotic behavior of kink solutions (exponential or power-law), as well as vibrational mode(s) can be present or absent in the kink's excitation spectrum. Thus, the $\varphi^8$ model, on the one hand, is not too complicated, and on the other hand, the variety of topological solitons generated by it is pretty significant.

As already mentioned, the $\varphi^8$ model can have kink solutions with one or both power-law tails; for more details, see, e.g., \cite[Sec.~II.A]{Christov.PRD.2019}. Such kinks have interesting features in comparison with their long-studied counterparts, which have exponential asymptotics. In particular, the forces of kink-kink and kink-antikink interactions decrease with distance much more slowly \cite{Radomskiy.JPCS.2017,Christov.PRL.2019,Manton.JPA.2019,Khare.JPA.2019,dOrnellas.JPC.2020,Campos.PLB.2021}. Note that the numerical treatment of kinks with power-law asymptotics is much more complicated and requires the use of various methods to suppress unwanted effects from overlapping power-law tails \cite[Sec.~III]{Christov.PRD.2019}, \cite[Sec.~2 and 3]{Christov.CNSNS.2021}.

Interestingly, the variety of asymptotic behaviors of kinks is not limited to exponential and power-law ones. In some models, kink solutions have super-exponential \cite{Kumar.IJMPB.2021,Khare.PS.2019} and power-tower \cite{Khare.JPA.2019.log} asymptotics. In addition, it is possible to construct so-called compact kinks that have no tails at all, i.e., have a compact support, see, e.g., \cite{Bazeia.PLB.2014,Bazeia.EPL.2014} and references therein.

Returning to the scattering of kinks in (1+1)-dimensional field-theoretic models, we mention the formation of oscillons --- moving or static oscillating structures localized in space, which also can form bound states. Escaping oscillons, as well as bound states of oscillons, have been observed, e.g., in the sinh-deformed $\varphi^4$ and the double sine-Gordon models \cite{Bazeia.EPJC.2018.sinh,Gani.EPJC.2018,Gani.EPJC.2019,Simas.JHEP.2020}. Formation of oscillons in kink-antikink collisions was also found in hyperbolic models \cite{Bazeia.PLB.2021} and in a parametrized $\varphi^4$ model \cite{Nzoupe.MPLA.2021}. In \cite{Romanczukiewicz.JHEP.2018} a relation between the oscillational normal modes and the oscillon configurations was demonstrated. Production of a kink-antikink pair in the collision of wave trains was observed in \cite{Romanczukiewicz.PRL.2010}. Interestingly, the oscillon played the role of an intermediate state in this process. Kink-antikink pair creation, as a result of the excitation of the vibrational mode of the $\varphi^4$ kink, was observed in \cite{Romanczukiewicz.JPA.2006}. Production of antikink-kink pairs and solitary oscillating structures was also found in the double sine-Gordon model \cite{Simas.JHEP.2020}.

This paper focuses on studying various processes in collisions of two or more kinks of the $\varphi^8$ model. First, the transition of a kink from one topological sector to another, i.e., the annihilation of a kink belonging to one topological sector and creation of a new kink in another topological sector. Second, the formation of static or moving oscillons. Third, the production of kink-antikink pairs resulting from the collision of oscillons, which, in turn, are the product of the annihilation of the first generation of kinks and antikinks.

The paper is organized as follows. In Section \ref{sec:Model}, we introduce the model, describe notations, and give some necessary comments on the numerical methods and initial conditions. Sections \ref{sec:Results} and \ref{sec:K-AK} present our main results. Most of the results are visualized in figures and summarized in three tables for the reader's convenience. Finally, we conclude in Section \ref{sec:Conclusion}.

\section{Model and method}
\label{sec:Model}

\begin{figure*}[t!]
\centering
 \subfigure[]{\includegraphics[width=0.40
\textwidth, height=0.20 \textheight]{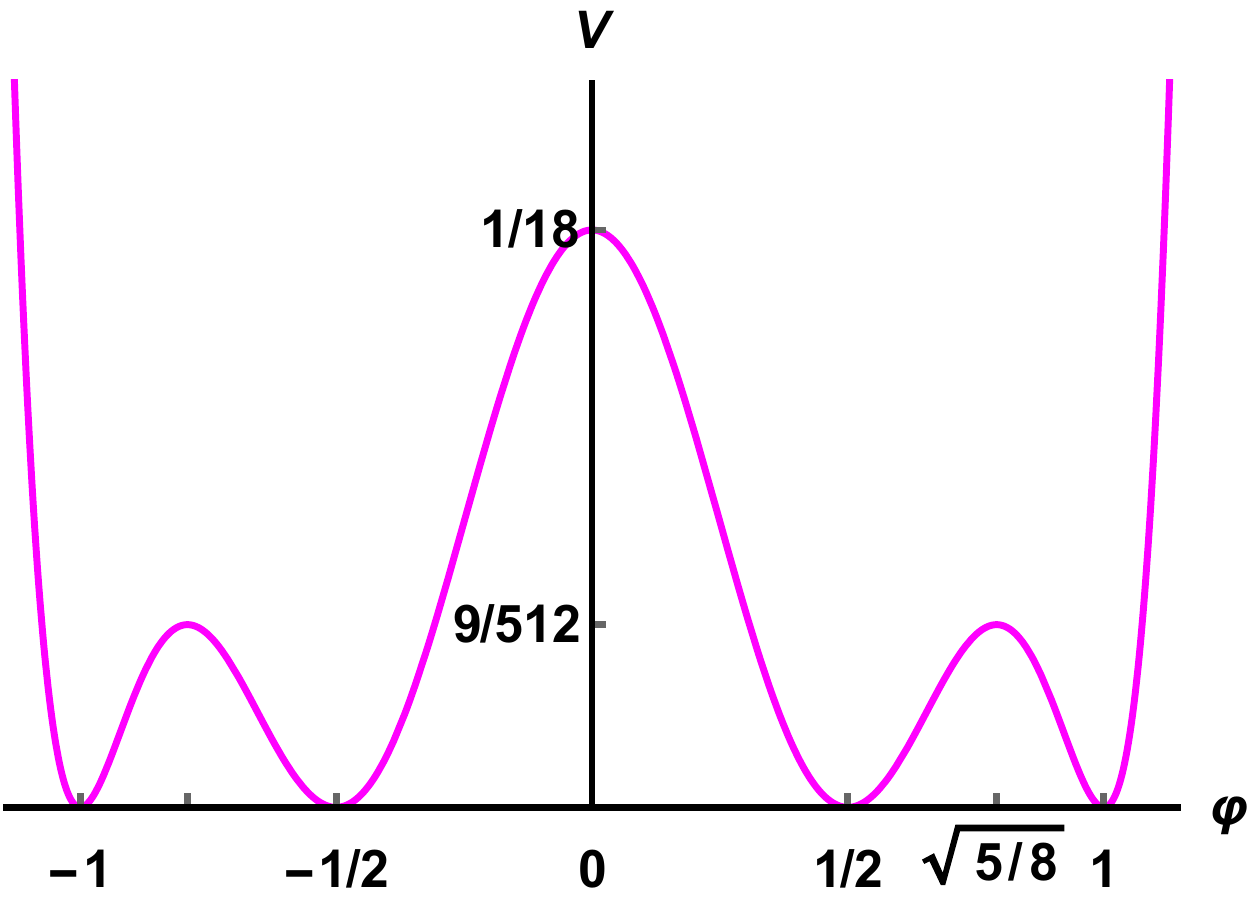}\label{fig:Phi8Potential_4minima}}
 \hspace{1mm}
 \subfigure[]{\includegraphics[width=0.40 \textwidth, height=0.20 \textheight]{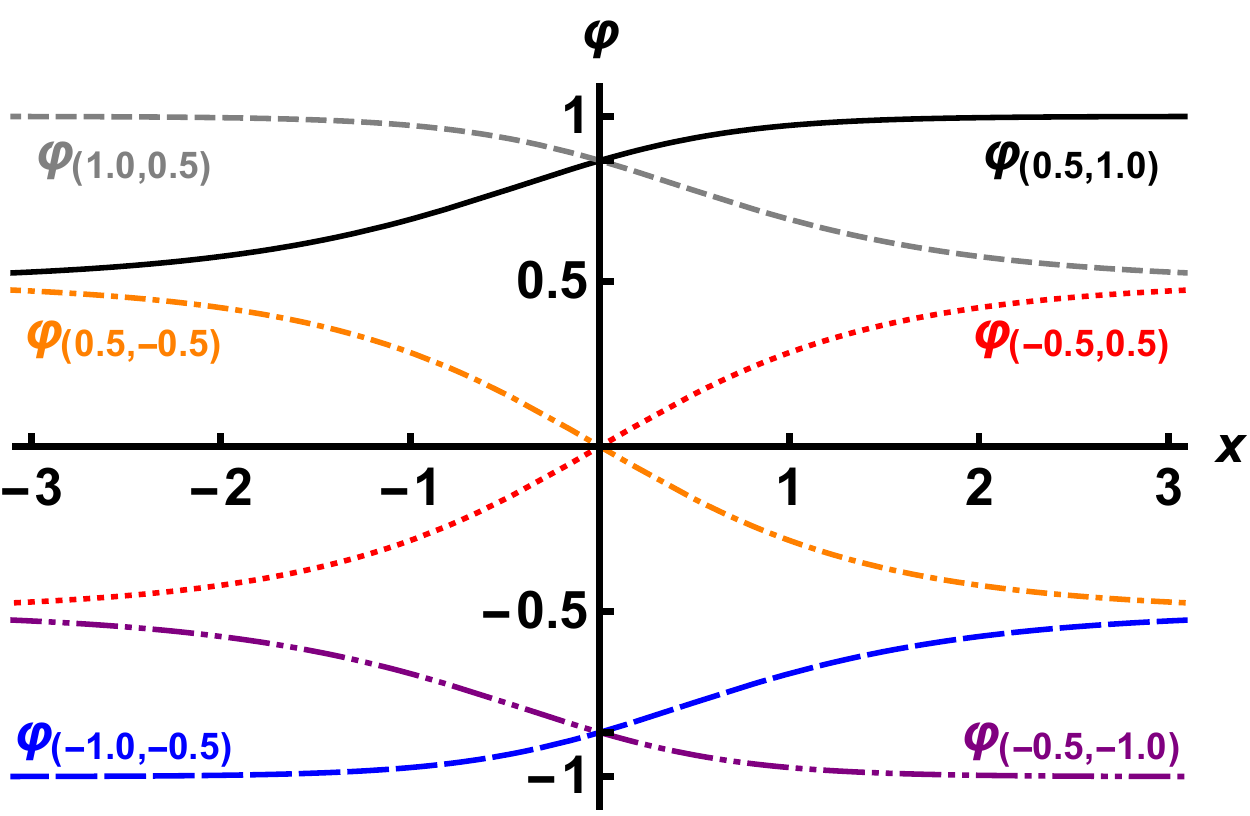}\label{fig:Phi8Solutions}}
 \hspace{1mm}
\caption{(a) The potential \eqref{eq:potential} of the $\varphi^8$ model. (b) Kinks \eqref{eq:kinks} of the $\varphi^8$ model.}
\label{fig:Phi8SolutionsPotential}
\end{figure*}

In its most general form, the potential of the $\varphi^8$ model is an eighth-degree polynomial with several degenerate minima (for convenience, we assume that the value of the potential at the minimum points is zero). As mentioned in the Introduction, some kinds of the $\varphi^8$ model have already been studied in the literature. In this paper, we consider a particular case
described by the Lagrangian
\begin{equation}\label{eq:Largangian}
\mathcal{L} = \frac{1}{2}\varphi_t^2 - \frac{1}{2}\varphi_x^2 - V(\varphi)
\end{equation}
with the potential term
%
\begin{equation}\label{eq:potential}
V(\varphi) = \frac{8}{9} \left(\varphi^2-1\right)^2 \left(\varphi^2-\frac{1}{4}\right)^2,
\end{equation}
see Fig.~\ref{fig:Phi8Potential_4minima}. Such a choice of potential means that there are four vacua, $\varphi=\pm 1$ and $\varphi=\pm\frac{1}{2}$, at which the potential vanishes. Thus, the model allows the existence of kink solutions in three topological sectors: $(-1,-\frac{1}{2})$, $(-\frac{1}{2},\frac{1}{2})$, and $(\frac{1}{2},1)$.

The Lagrangian \eqref{eq:Largangian} yields the equation of motion --- partial differential equation
\begin{equation}\label{eq:EOM}
\varphi_{tt}^{} - \varphi_{xx}^{} + V^\prime(\varphi) = 0.
\end{equation}
The energy functional for the Lagrangian \eqref{eq:Largangian} is
\begin{equation}\label{eq:energy}
E[\varphi] = \int\limits_{-\infty}^{+\infty} \left[ \frac{1}{2}\varphi_t^2 + \frac{1}{2}\varphi_x^2 + V(\varphi) \right]dx.   
\end{equation}
In the static case, the equation of motion \eqref{eq:EOM} can be reduced to the first-order ordinary differential equation
\begin{equation}\label{eq:Bogomolny}
\varphi_x^{} = \pm \sqrt{2V(\varphi)}.
\end{equation}
There are two kink solutions of Eq.~\eqref{eq:Bogomolny} in each topological sector: one is the strictly monotonic increasing function of $x$, called {\it kink}, and the other is the strictly monotonic decreasing function of $x$, called {\it antikink}. All three kinks and three antikinks of the model can be described by a single formula:
\begin{equation}\label{eq:kinks}
\varphi(x) = \cos \left(\frac{1}{3} \arccos(\tanh x)+\frac{\pi}{3}l\right),
\end{equation}
where the parameter $l$ takes any six consecutive integer values, at which all six topological solitons of the model are obtained. For example, at $l=0$, $1$, and $2$, Eq.~\eqref{eq:kinks} gives kinks in the sectors $(\frac{1}{2},1)$, $(-\frac{1}{2},\frac{1}{2})$, and $(-1,-\frac{1}{2})$, respectively; while at $l=3$, $4$, and $5$, Eq.~\eqref{eq:kinks} gives antikinks in the sectors $(-\frac{1}{2},-1)$, $(\frac{1}{2}, -\frac{1}{2})$, and $(1,\frac{1}{2})$, respectively; see Fig.~\ref{fig:Phi8Solutions}.

The energy of the static kink can be found by substituting Eq.~\eqref{eq:kinks} into Eq.~\eqref{eq:energy}, which is called kink's {\it mass}. Masses of the kink and antikink in a given topological sector are obviously the same. (Note that it is easy to show that the kink masses in all topological sectors can also be found without finding the kinks themselves, see, e.g., \cite[Sec.~II]{Gani.PRD.2020}.)

In the case of the potential \eqref{eq:potential}, there are two different types of kink solutions: symmetric kinks in the sector $(-\frac{1}{2},\frac{1}{2})$ and asymmetric kinks in the sectors $(-1,-\frac{1}{2})$ and $(\frac{1}{2},1)$. Their masses are the following:
\begin{equation}\label{eq:masses}
M_{\rm K}^{} =\frac{19}{90} \quad \mbox{and} \quad M_{\rm k}^{} = \frac{11}{180},
\end{equation}
for symmetric and asymmetric kinks, respectively. Note that the symmetric kink is about $3.5$ times heavier than the asymmetric one. At the same time, we found that in the excitation spectra of both kinks, there is only the zero (translational) mode and no vibrational modes. (For details of the kink's excitation spectrum search, see, e.g., \cite[Sec.~IV]{Gani.PRD.2020}.)

Let us also agree on the notation used further in the text. We will symbolically denote the symmetric kinks by a capital ``K'' and the asymmetric ones by a lowercase ``k'' (this notation was already used in Eq.~\eqref{eq:masses}). In addition, in the final states of kink collisions, we will encounter objects that can be classified as {\it bion} (will be denoted by ``b'') --- a bound state of a kink and an antikink, and {\it oscillon} (will be denoted by ``o'') --- a localized formation oscillating about the vacuum with a small amplitude. The smallness of the amplitude of field oscillations near the vacuum value means that the amplitude is significantly less than the distance to the neighboring vacuum. In this sense, oscillon differs from a bion, in which the field oscillates with an amplitude close to the distance to the neighboring vacuum of the model. Besides that, in most cases in the final state, there is radiation in the form of small-amplitude waves, which carries away energy from the collision region. We will also use the term {\it type of configuration} in the following sense. For example, in the case of collision of the kink and antikink belonging to the topological sector $(-\frac{1}{2},\frac{1}{2})$, in the initial state, we have a configuration, which we will call a configuration of the type $(-\frac{1}{2},\frac{1}{2},-\frac{1}{2})$, and so on in the same spirit.

We performed the numerical simulation of scattering of two, three, and four solitary waves (kinks and antikinks) in different topological sectors. The equation of motion \eqref{eq:EOM} was solved numerically using discretization of the fourth order in space and the St{\"o}rmer method of integration with respect to time, with the spatial and temporal steps being equal to 0.025 and 0.005, respectively.

The initial conditions in all cases were constructed in the form of two, three, or four kinks (antikinks) $\varphi_{s_i^{}}^{}$ located at the points $X_i^{}$ and moving with the velocities $v_i^{}$:
\begin{equation}\label{eq:initial_condition}
    \varphi(x,t) = \sum_{i=1}^n\varphi_{s_i^{}}^{}\left(\frac{x-X_i^{}-v_i^{}t}{\sqrt{1-v_i^2}}\right) + C_n^{},
\end{equation}
with $n=2$, 3, and 4 in collisions of two, three, and four solitons, respectively. The subscript $s_i^{}$ stands for the topological sector of the $i$-th kink/antikink, and can take ``values'' $(-1,-\frac{1}{2})$, $(-\frac{1}{2},-1)$, $(-\frac{1}{2},\frac{1}{2})$, $(\frac{1}{2},-\frac{1}{2})$, $(\frac{1}{2},1)$, or $(1,\frac{1}{2})$. The constants $C_n^{}$ in Eq.~\eqref{eq:initial_condition} are chosen so that at $x\to\pm\infty$ and in between the kinks the field tends to the vacuum values.

\section{Results: Zoo of the final states in the collisions of kinks}
\label{sec:Results}

\subsection{Two kinks}
\label{sec:Results.Two}

We have performed numerical simulations of the following two-kink processes.
\begin{enumerate}

    \item Collision of symmetric kink and antikink belonging to the topological sector $(-\frac{1}{2},\frac{1}{2})$. We used the initial configuration of the type $(-\frac{1}{2},\frac{1}{2},-\frac{1}{2})$, given by Eq.~\eqref{eq:initial_condition} with $n=2$, $s_1^{}=(-\frac{1}{2},\frac{1}{2})$, $s_2^{}=(\frac{1}{2},-\frac{1}{2})$, $C_2^{}=-\frac{1}{2}$, $-X_1^{}=X_2^{}=10$, $v_1^{}=-v_2^{}=0.1$. The picture of the collision is shown in Fig.~\ref{fig:m0505m05collision}. The kink and the antikink collide at the origin, annihilate each other and, forming two oscillons escaping from the collision point at high velocities $v_{\rm o_{1,2}^{}}=\pm 0.67$. Hence, the observed reaction could be symbolically written as $K+\bar{K}\to o+o$. The masses of the produced oscillons are found to be about $m_{\rm o}^{}=0.193$. In addition, part of the energy is emitted in the form of small-amplitude waves.
\begin{figure}[t!]
\begin{center}
  \centering
    \subfigure[\:$(-\frac{1}{2},\frac{1}{2},-\frac{1}{2})$]{\includegraphics[width=0.40
 \textwidth]
{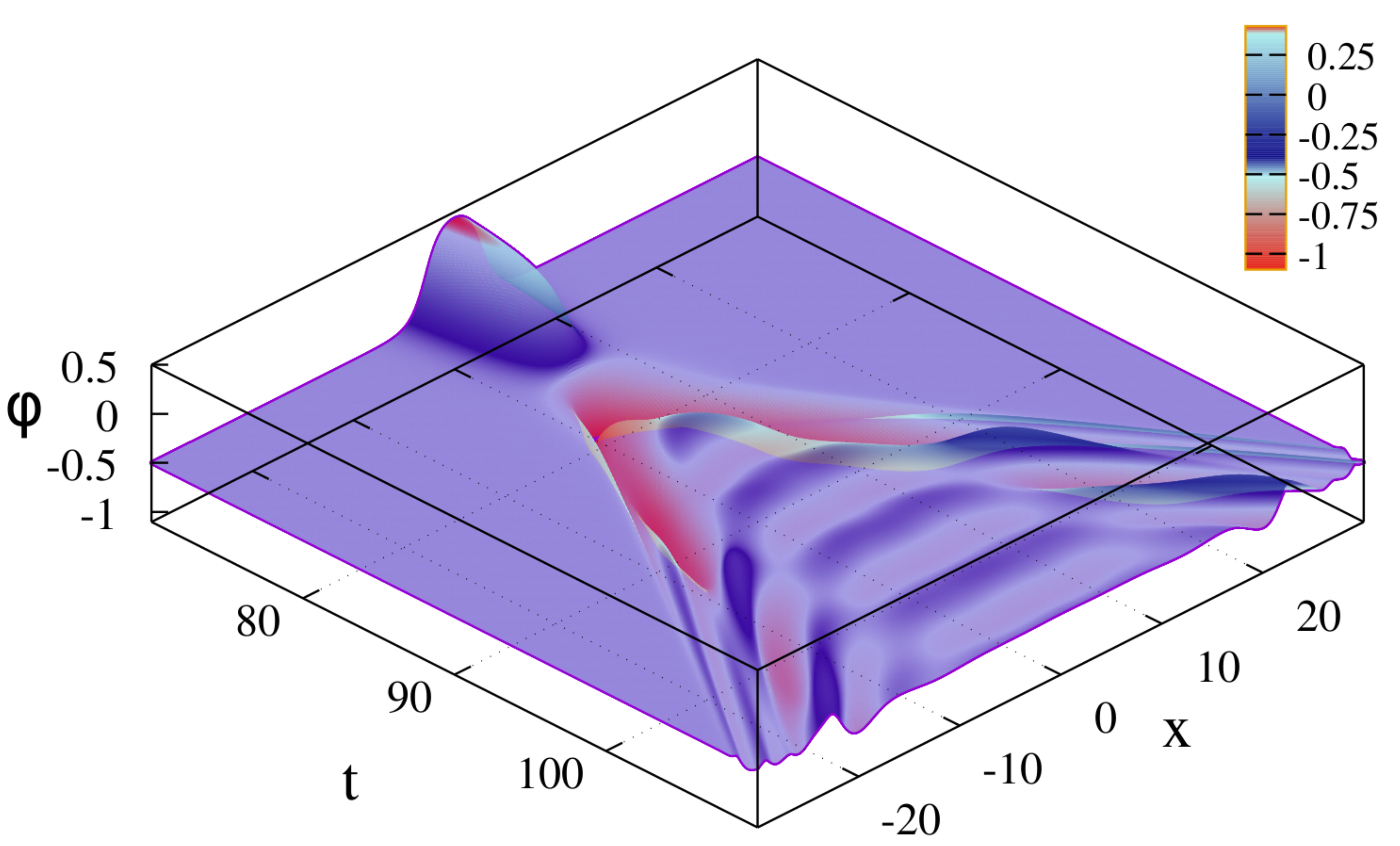}\label{fig:m0505m05collision}}
  \subfigure[\:$(\frac{1}{2},1,\frac{1}{2})$]{\includegraphics[width=0.40
 \textwidth]
{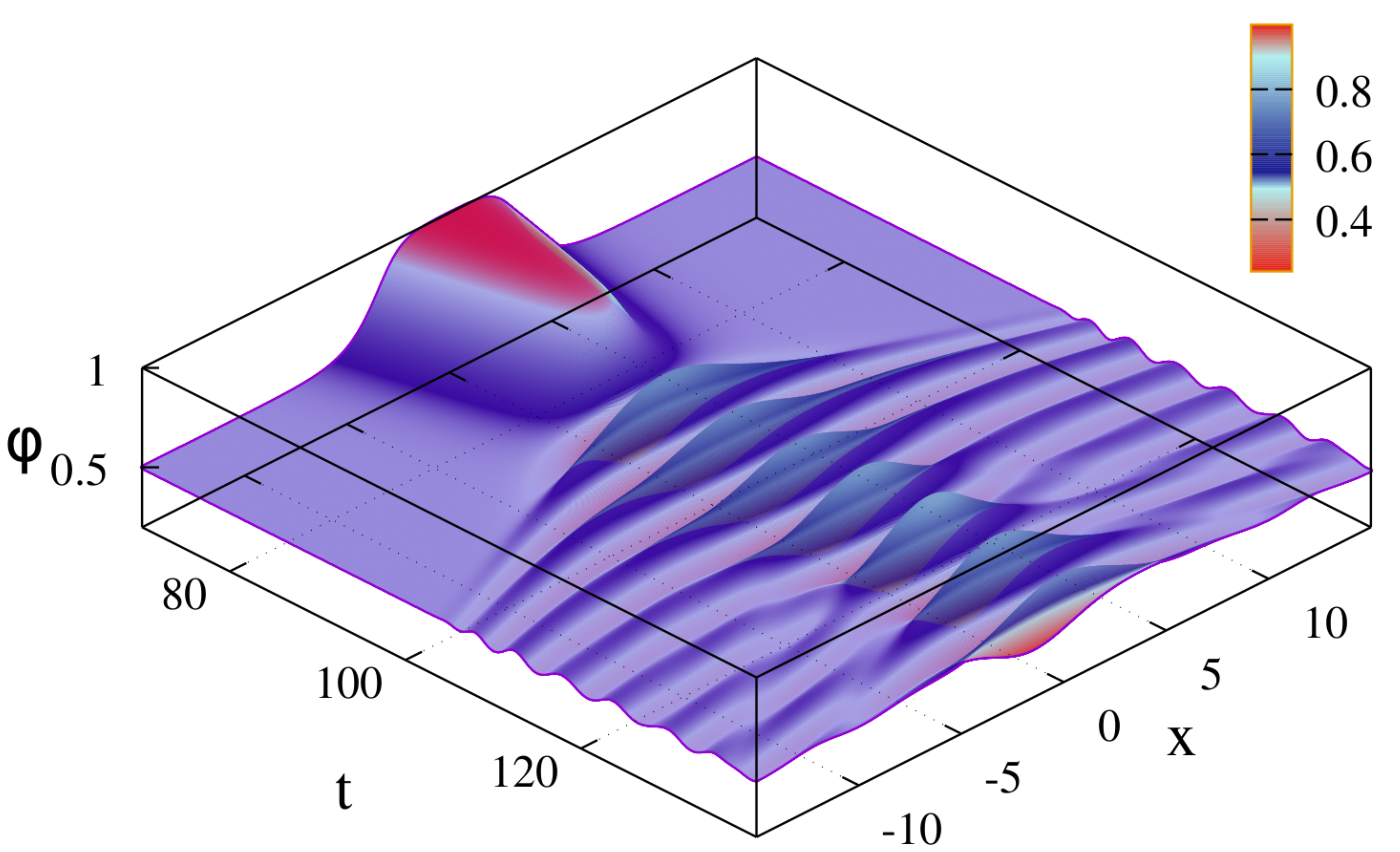}\label{fig:051005collision}}
\\
  \subfigure[\:$(1,\frac{1}{2},1)$]{\includegraphics[width=0.40
 \textwidth]
{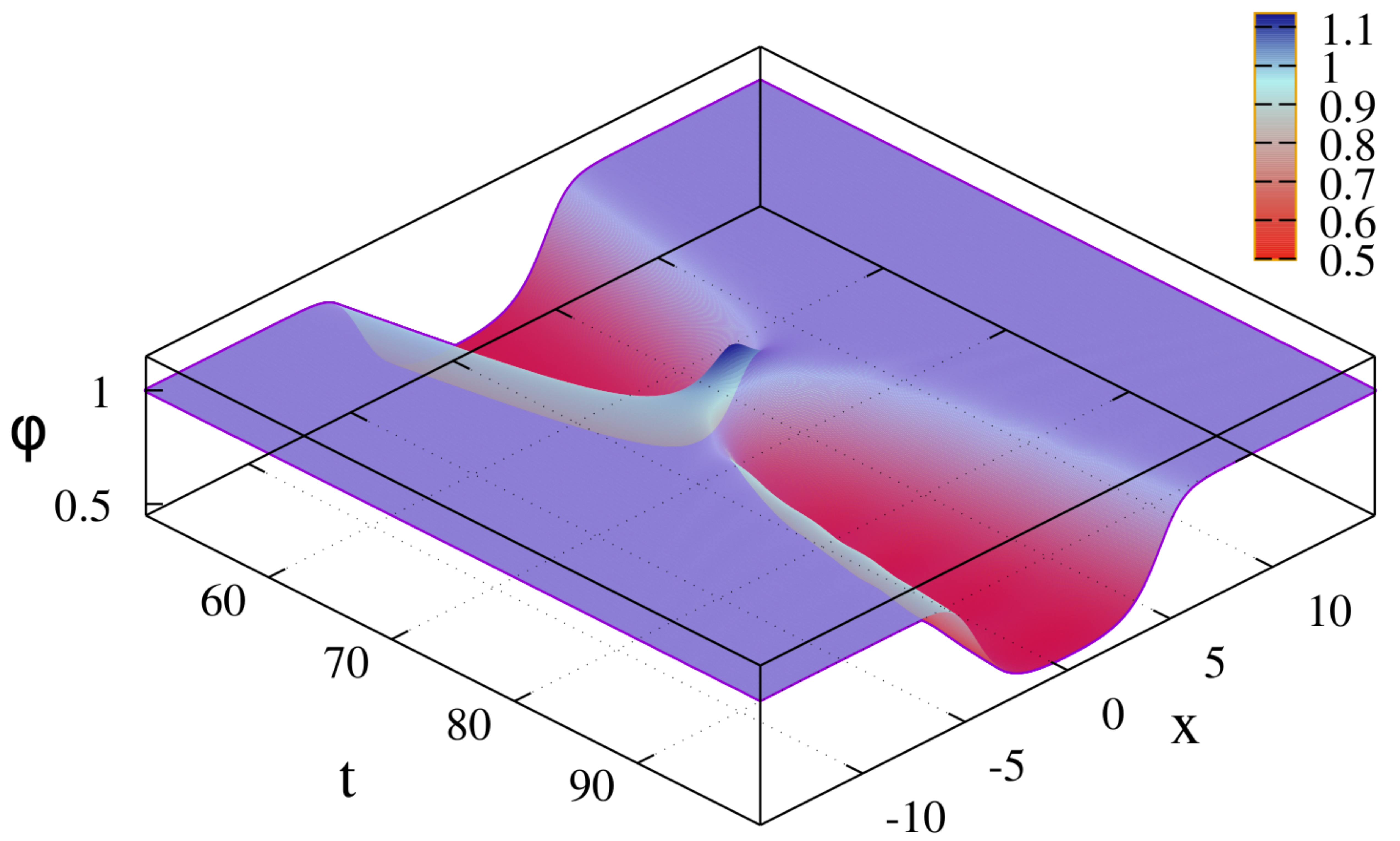}\label{fig:100510collision}}
      \subfigure[\:$(-1,-\frac{1}{2},\frac{1}{2})$]{\includegraphics[width=0.40
 \textwidth]
{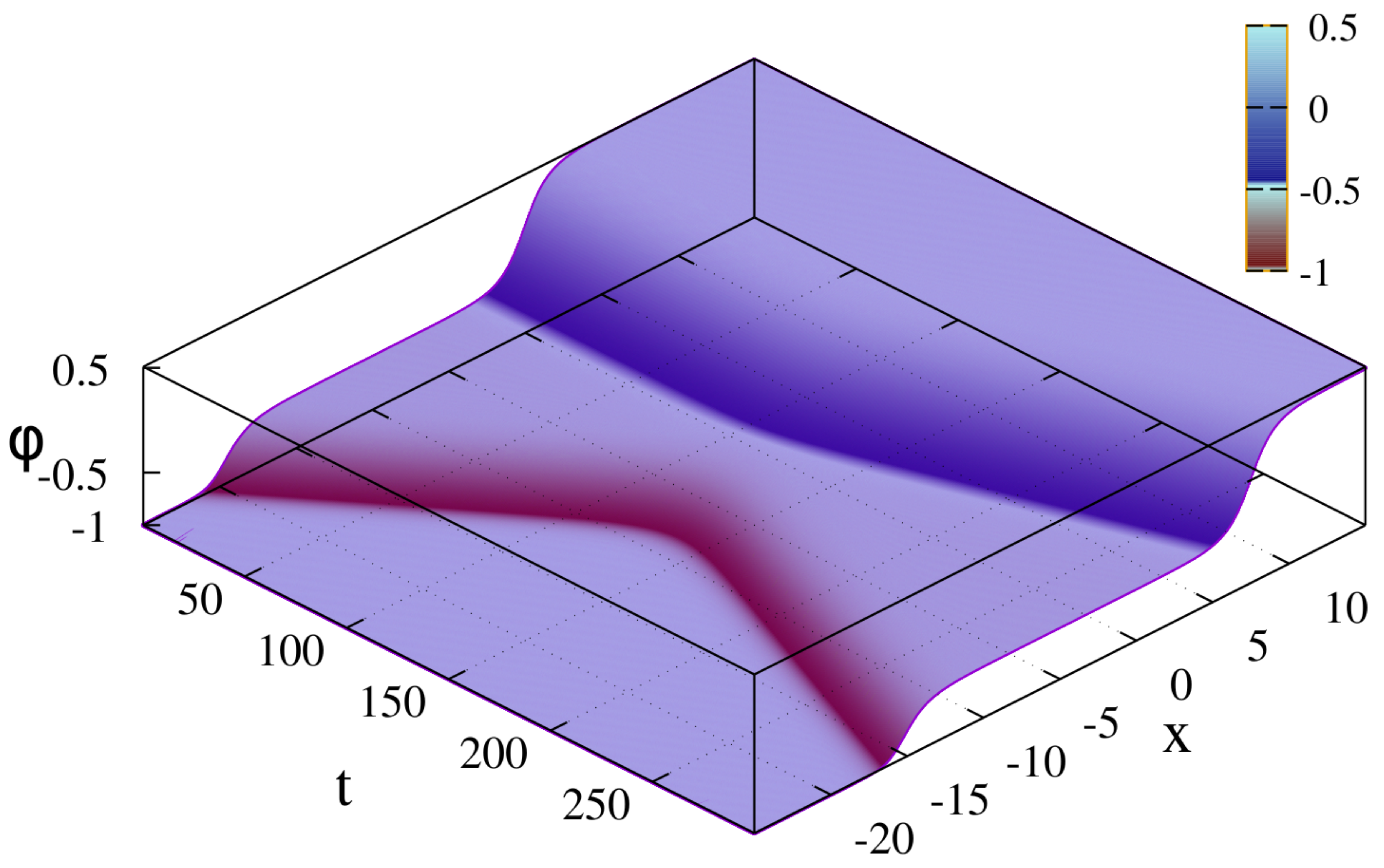}\label{fig:V01V00Xm20X00Fieldm10m0505collision}}
  \caption{Space-time plots of two-kink collisions.}
  \label{fig:2kcollision}
\end{center}
\end{figure}

    \item Collision of asymmetric kink and antikink belonging to the topological sector $(\frac{1}{2},1)$. We used the initial configuration of the type $(\frac{1}{2},1,\frac{1}{2})$, given by Eq.~\eqref{eq:initial_condition} with $n=2$, $s_1^{}=(\frac{1}{2},1)$, $s_2^{}=(1,\frac{1}{2})$, $C_2^{}=-1$, $-X_1^{}=X_2^{}=10$, $v_1^{}=-v_2^{}=0.1$. In this case, the initial velocities of the kinks are less than the critical value $v_{\rm cr}\approx 0.88$, therefore, the kink and antikink become trapped and their bound state (bion) is formed; see Fig.~\ref{fig:051005collision}. In this case, the observed reaction is $k+\bar{k}\to b$.

    \item Collision of asymmetric antikink and kink belonging to the topological sector $(\frac{1}{2},1)$. We used the initial configuration of the type $(1,\frac{1}{2},1)$, given by Eq.~\eqref{eq:initial_condition} with $n=2$, $s_1^{}=(1,\frac{1}{2})$, $s_2^{}=(\frac{1}{2},1)$, $C_2^{}=-\frac{1}{2}$, $-X_1^{}=X_2^{}=10$, $v_1^{}=-v_2^{}=0.1$. The initial velocities are now higher than $v_{\rm cr}\approx 0.07777$, therefore, the antikink and kink collide once and escape to spatial infinities; see Fig.~\ref{fig:100510collision}. The final velocities of the kinks are $-v_{\rm 1f}^{}=v_{\rm 2f}^{}=0.064$. Hence, in this case we observe inelastic scattering $\bar{k}+k\to \bar{k}+k$.

    \item Collision of asymmetric kink $(-1,-\frac{1}{2})$ and symmetric kink $(-\frac{1}{2},\frac{1}{2})$.

    First, we used the initial configuration of the type $(-1,-\frac{1}{2},\frac{1}{2})$, given by Eq.~\eqref{eq:initial_condition} with $n=2$, $s_1^{}=(-1,-\frac{1}{2})$, $s_2^{}=(-\frac{1}{2},\frac{1}{2})$, $C_2^{}=\frac{1}{2}$, $X_1^{}=-20$, $X_2^{}=0$, $v_1^{}=0.1$, $v_2^{}=0$. This means that the light asymmetric kink approaches the heavy symmetric kink, which is initially at rest at the origin, see Fig.~\ref{fig:V01V00Xm20X00Fieldm10m0505collision}. The final velocity of the light kink after the collision is $v_{\rm 1f}^{}=-0.055$, while the heavy kink has been scattered with the final velocity $v_{\rm 2f}^{}=0.045$.

    Second, we modified the initial conditions such that the heavy symmetric kink had the initial velocity $v_{2}^{}=-0.1$ and position $X_2^{}=20$, while the light asymmetric kink was initially at rest at the origin.
    The final velocities of the light and heavy kinks are $v_{\rm 1f}^{}=-0.154$ and $v_{\rm 2f}^{}=-0.055$, respectively.

    It seems that no radiation of energy in the form of small-amplitude waves is observed for the two above scenarios. Therefore, we can say that we observe elastic scattering $k+K\to k+K$ in this case.

\end{enumerate}
For the convenience of the reader, we summarize the above information in Table \ref{tab:Table_1}.

Here we have shown only examples of typical two-kink processes. A more detailed analysis of kink-antikink and antikink-kink collisions in all topological sectors will be carried out below in Section \ref{sec:K-AK}.

\begin{table}[t!]
\begin{ruledtabular}
\caption{Summary of processes observed in the collisions of two kinks/antikinks.}
\label{tab:Table_1}
\begin{tabular}{ccccc}
type & ``reaction'' & initial
velocities & final velocities & figure\\
\hline
\vphantom{$\displaystyle\frac{1}{2}$} $(-\frac{1}{2},\frac{1}{2},-\frac{1}{2})$ & $K\bar{K}\to oo$ & 
$v_1^{}=-v_2^{}=0.1$ & $v_{\rm o_{1,2}^{}}=\pm 0.67$ & \ref{fig:m0505m05collision}\\
\vphantom{$\displaystyle\frac{1}{2}$} $(\frac{1}{2},1,\frac{1}{2})$ & $k\bar{k}\to b$ & 
$v_1^{}=-v_2^{}=0.1$ & $v_{\rm b}^{}=0$ & \ref{fig:051005collision}\\
\vphantom{$\displaystyle\frac{1}{2}$} $(1,\frac{1}{2},1)$ & $\bar{k}k\to \bar{k}k$ & 
$v_1^{}=-v_2^{}=0.1$ & $-v_{\rm 1f}^{}=v_{\rm 2f}^{}=0.064$ & \ref{fig:100510collision}\\
\vphantom{$\displaystyle\frac{1}{2}$} $(-1,-\frac{1}{2},\frac{1}{2})$ & $kK\to kK$ & 
$v_1^{}=0.1$, $v_2^{}=0$ & $v_{\rm 1f}^{}=-0.055$, $v_{\rm 2f}^{}=0.045$ & \ref{fig:V01V00Xm20X00Fieldm10m0505collision}\\
\vphantom{$\displaystyle\frac{1}{2}$} $(-1,-\frac{1}{2},\frac{1}{2})$ & $kK\to kK$ & 
$v_1^{}=0$, $v_{2}^{}=-0.1$ & $v_{\rm 1f}^{}=-0.154$, $v_{\rm 2f}^{}=-0.055$ & 
--- \\
\end{tabular}
\end{ruledtabular}
\end{table}

\subsection{Three kinks}
\label{sec:Results.Three}

We have performed numerical simulations of the following three-kink processes.
\begin{enumerate}

    \item First of all, we studied the collision of two kinks and one antikink belonging to the topological sector $(-\frac{1}{2},\frac{1}{2})$, i.e.\ the kink-antikink-kink collision. For this purpose, we used the initial configuration of the type $(-\frac{1}{2},\frac{1}{2},-\frac{1}{2},\frac{1}{2})$. At $t=0$, the antikink was at rest at the origin, and two kinks were moving towards it with initial velocities 0.1 from the points $x=\pm 20$, see Fig.~\ref{fig:m0505m0505collision}. This corresponds to the initial configuration \eqref{eq:initial_condition} with $n=3$, $s_1^{}=s_3^{}=(-\frac{1}{2},\frac{1}{2})$, $s_2^{}=(\frac{1}{2},-\frac{1}{2})$, $C_3^{}=0$, $-X_1^{}=X_3^{}=20$, $X_2^{}=0$, $v_1^{}=-v_3^{}=0.1$, $v_2^{}=0$. All three solitary waves collide at the origin, forming an oscillating bound state for some time. This bound state looks like a bion formed by a kink and an antikink. At the same time, the kink, which started from the point $X_3^{}=20$, keeps its identity. The situation looks as if this kink was attracted to the bion, collided with it, bounced, then attracted again, and so on. Such oscillations occur several times. After that, the kink $(-\frac{1}{2},\frac{1}{2})$ escapes back in the direction of increasing $x$. Simultaneously, an exciting transformation occurs with the bion: a pair of a kink and an antikink, belonging to the topological sector $(-1,-\frac{1}{2})$ is formed in its place. These kinks, in turn, escape from each other at low relative velocity. Moreover, their center of mass moves with high speed in the direction of decreasing $x$. The above evolution can be clearly observed in Fig.~\ref{fig:m0505m0505collision}.

\begin{figure}[t!]
\begin{center}
  \centering
    \subfigure[\:$(-\frac{1}{2},\frac{1}{2},-\frac{1}{2},\frac{1}{2})$]{\includegraphics[width=0.32
 \textwidth]
{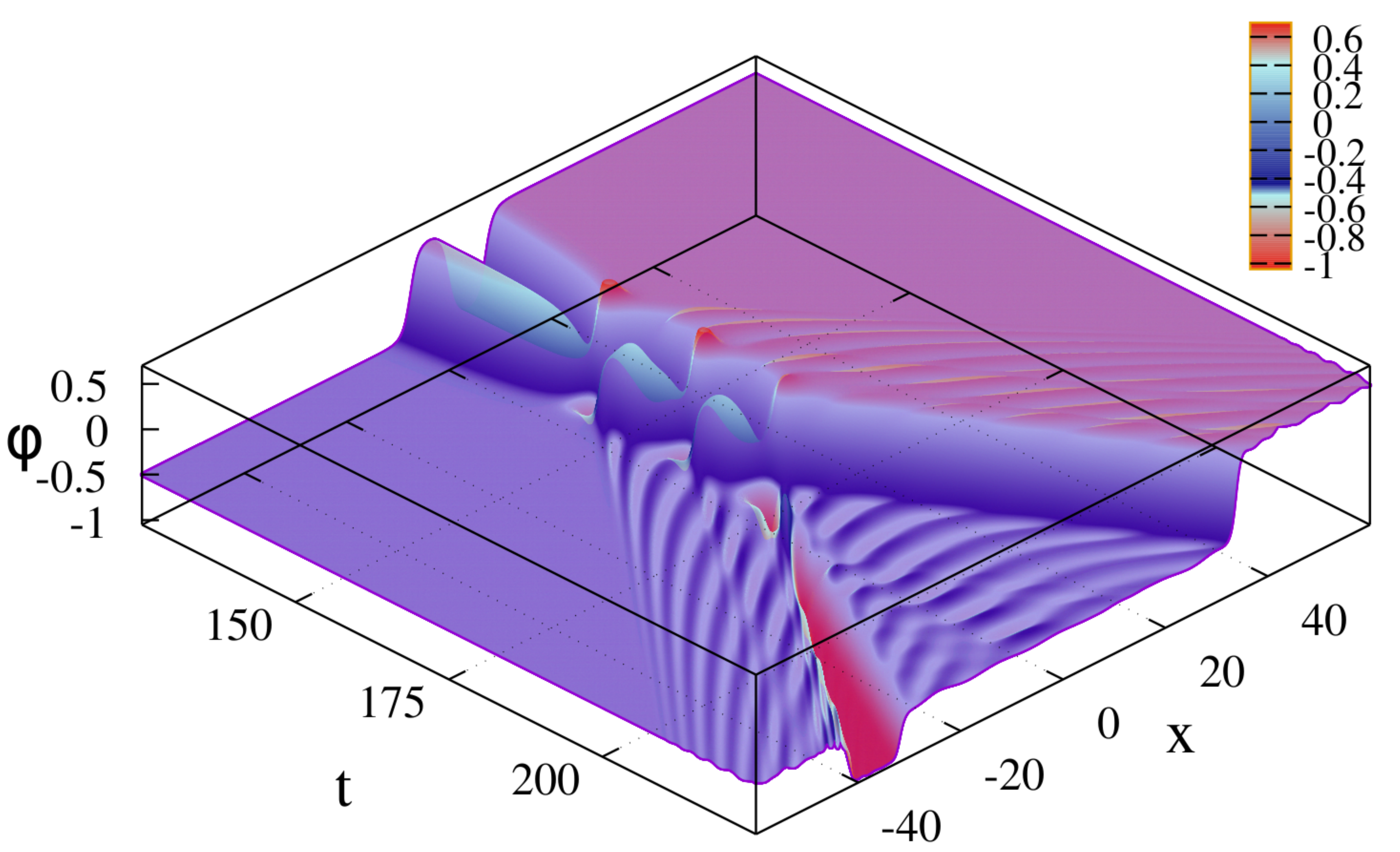}\label{fig:m0505m0505collision}}
    \subfigure[\:$(-\frac{1}{2},\frac{1}{2},-\frac{1}{2},\frac{1}{2})$]{\includegraphics[width=0.32
 \textwidth]
{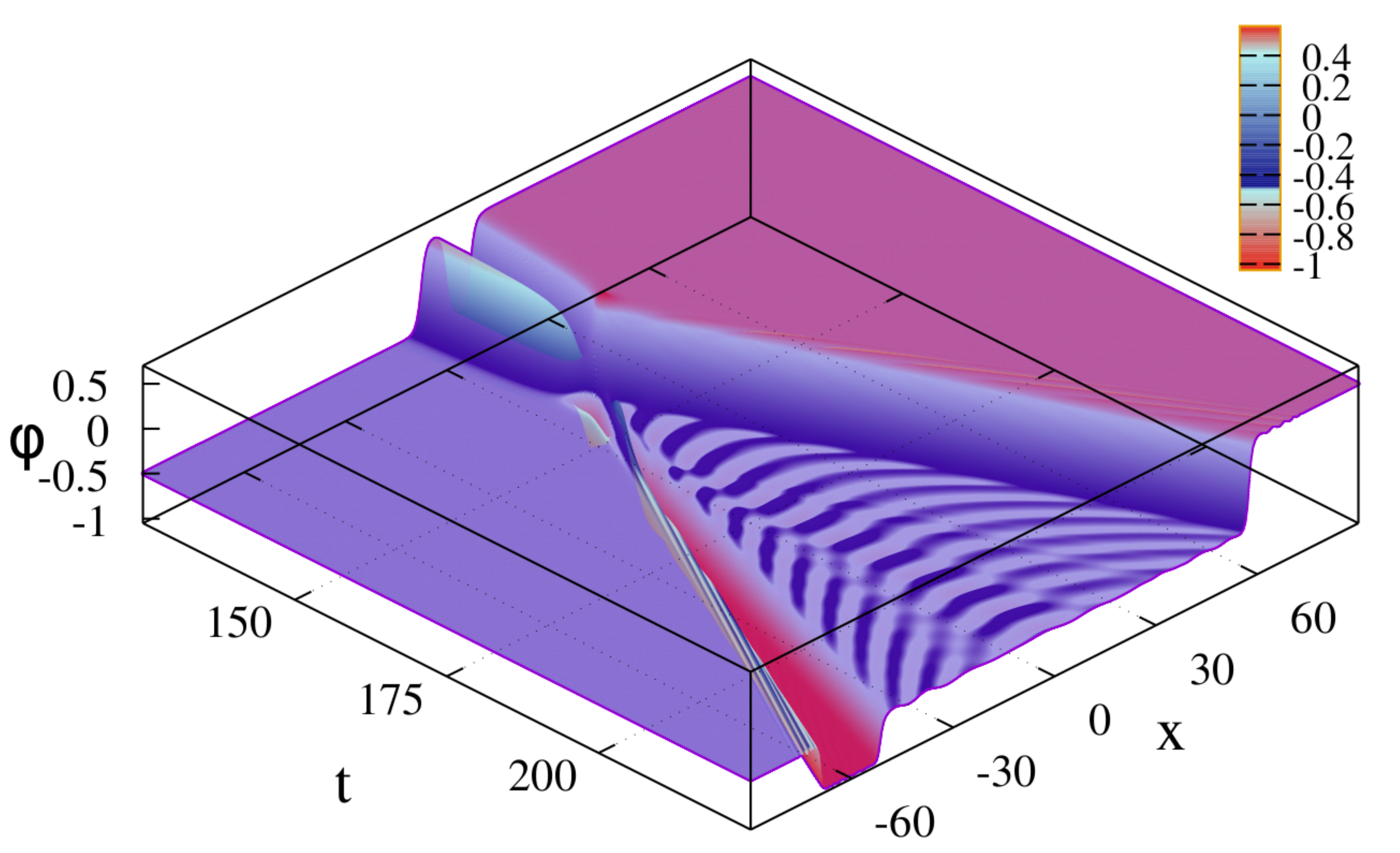}\label{fig:LKM1999}}
    \subfigure[\:$(-\frac{1}{2},\frac{1}{2},-\frac{1}{2},\frac{1}{2})$]{\includegraphics[width=0.32
 \textwidth]
{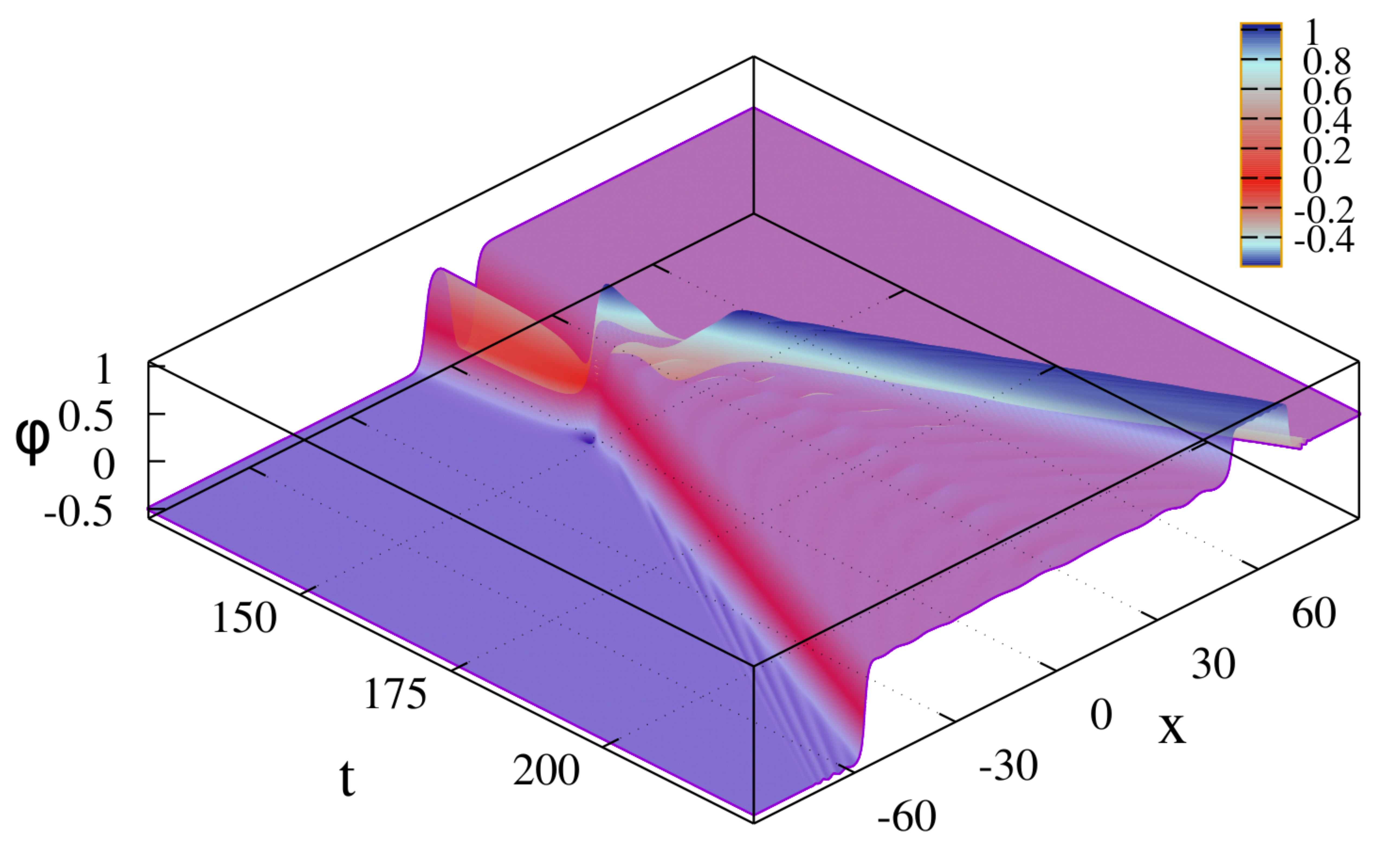}\label{fig:LKM2001}}
  \\
 \subfigure[\:$(-\frac{1}{2},\frac{1}{2},-\frac{1}{2},\frac{1}{2})$]{\includegraphics[width=0.32
 \textwidth]
{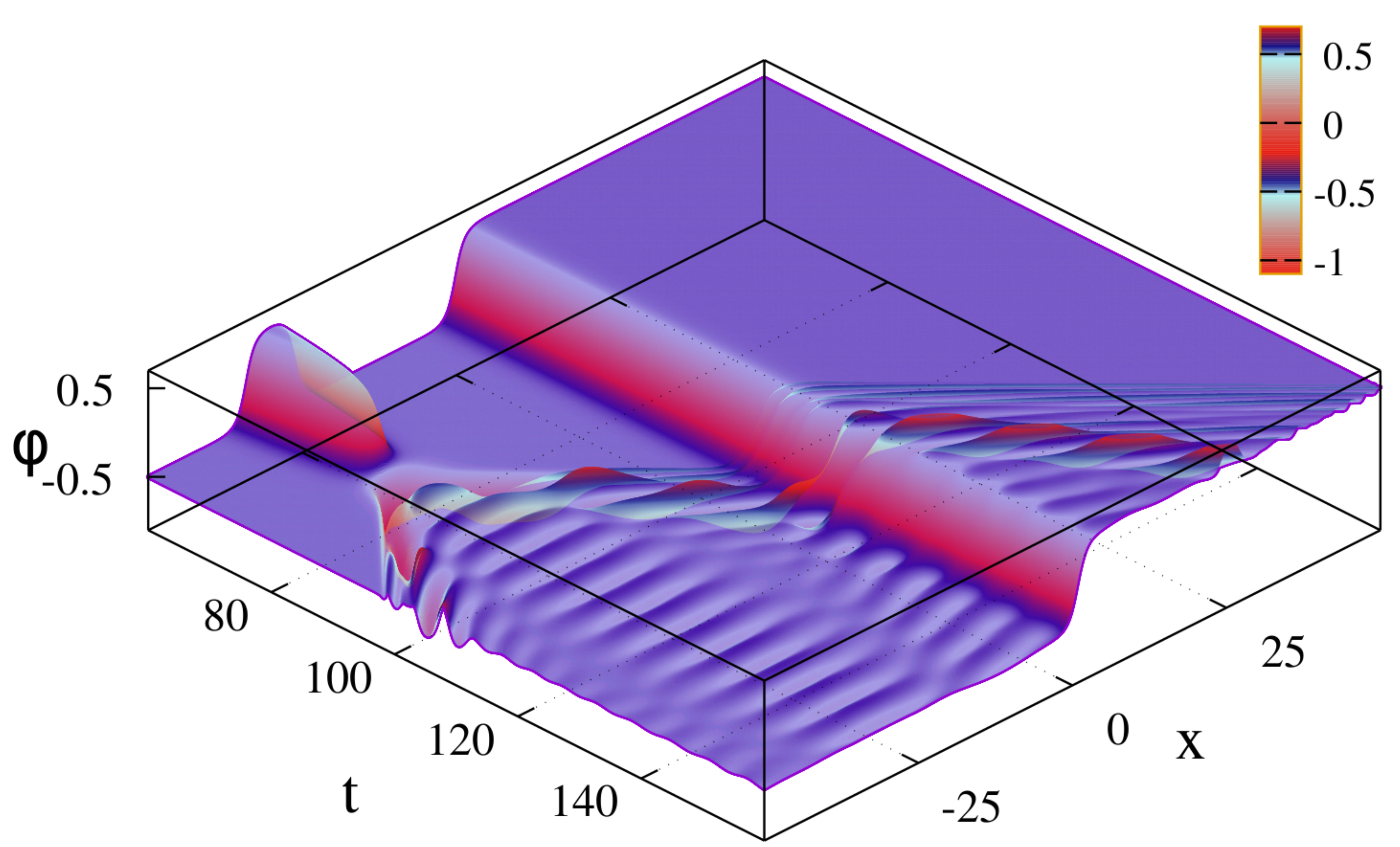}\label{fig:V01Vm01V00Xm10X10X30Fieldm05p05m05p05collision}}
\subfigure[\:$(\frac{1}{2},1,\frac{1}{2},1)$]{\includegraphics[width=0.32
 \textwidth]
{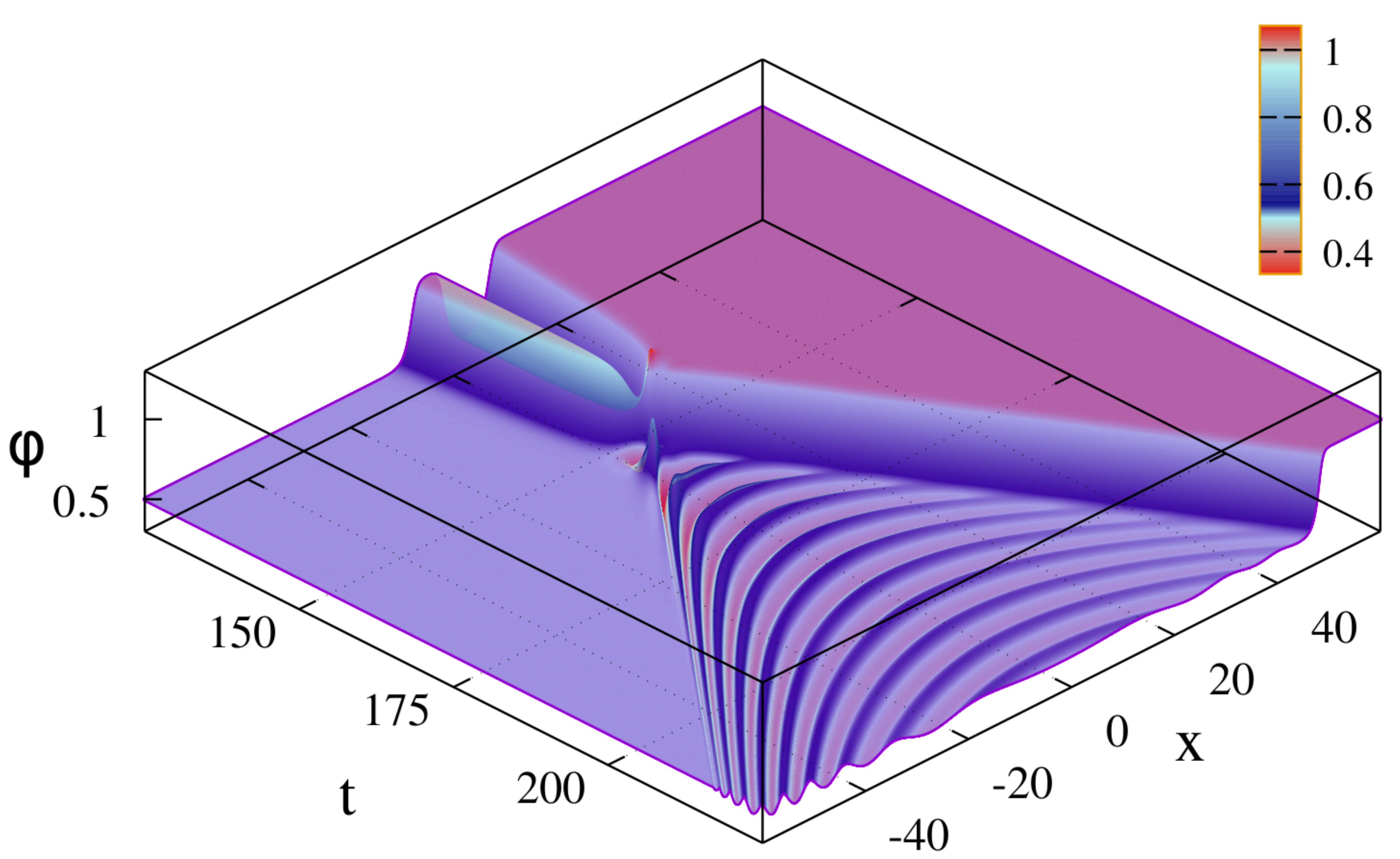}\label{fig:05100510collision}}
    \subfigure[\:$(\frac{1}{2},-\frac{1}{2},\frac{1}{2},1)$]{\includegraphics[width=0.32
 \textwidth]
{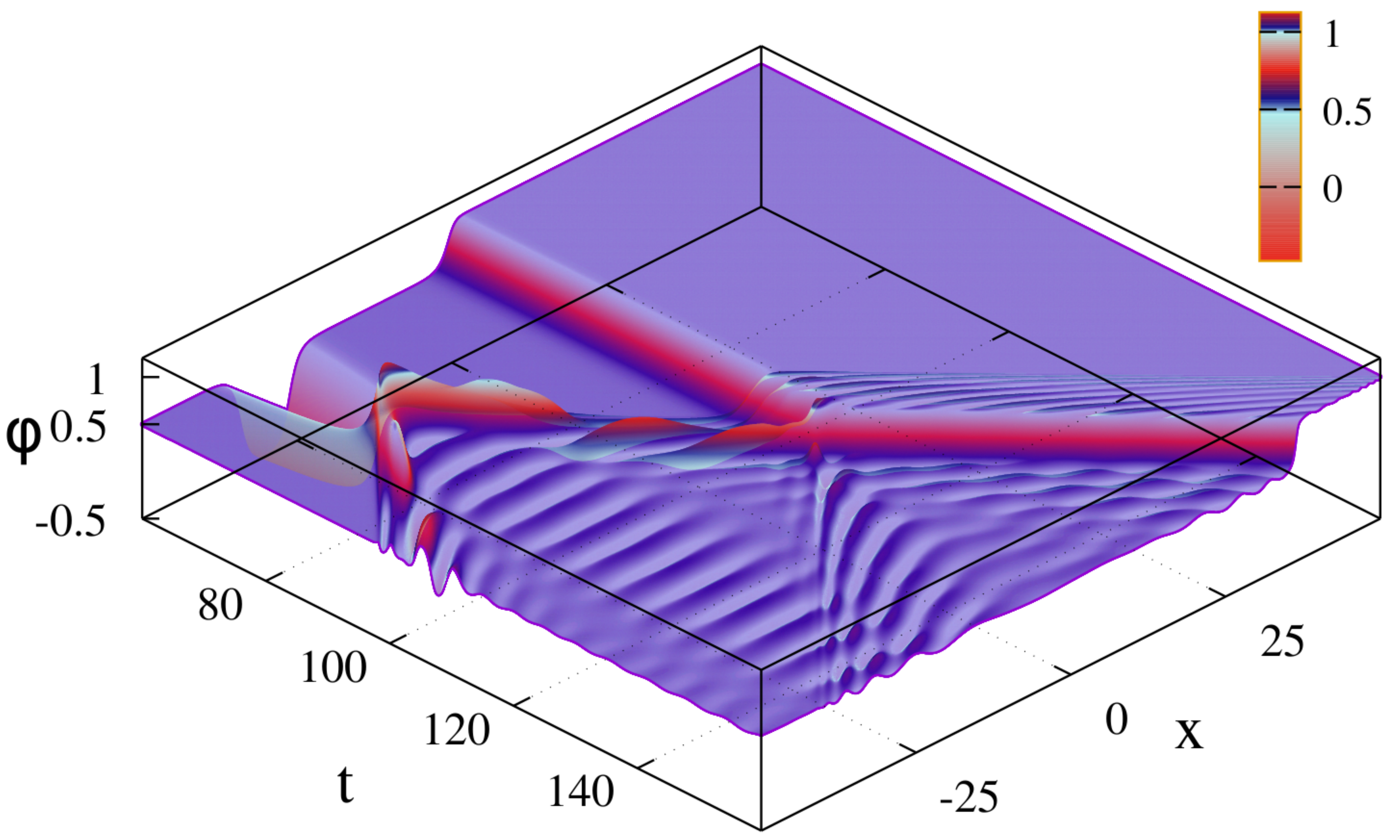}\label{fig:V01Vm01V00Xm10X10X30Fieldp05m05p05p10collision}}
  \\
  \subfigure[\:$(-\frac{1}{2},-1,-\frac{1}{2},\frac{1}{2})$]{\includegraphics[width=0.32
 \textwidth]
{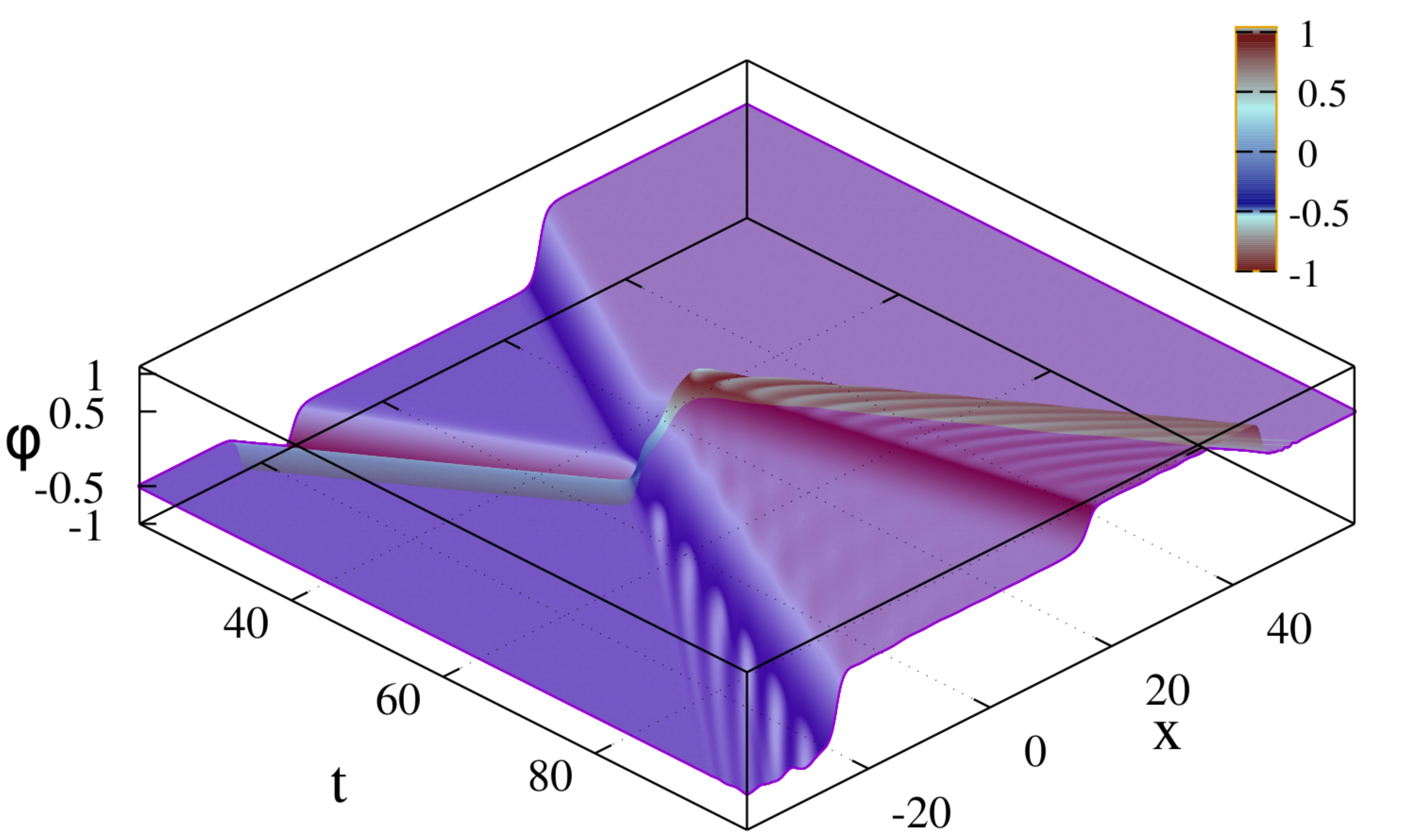}\label{fig:V084V065Vm07Xm400Xm280Xp400Fieldm05m10m05p05collision}}
   \subfigure[\:$(-1,-\frac{1}{2},\frac{1}{2},1)$]{\includegraphics[width=0.32
 \textwidth]
{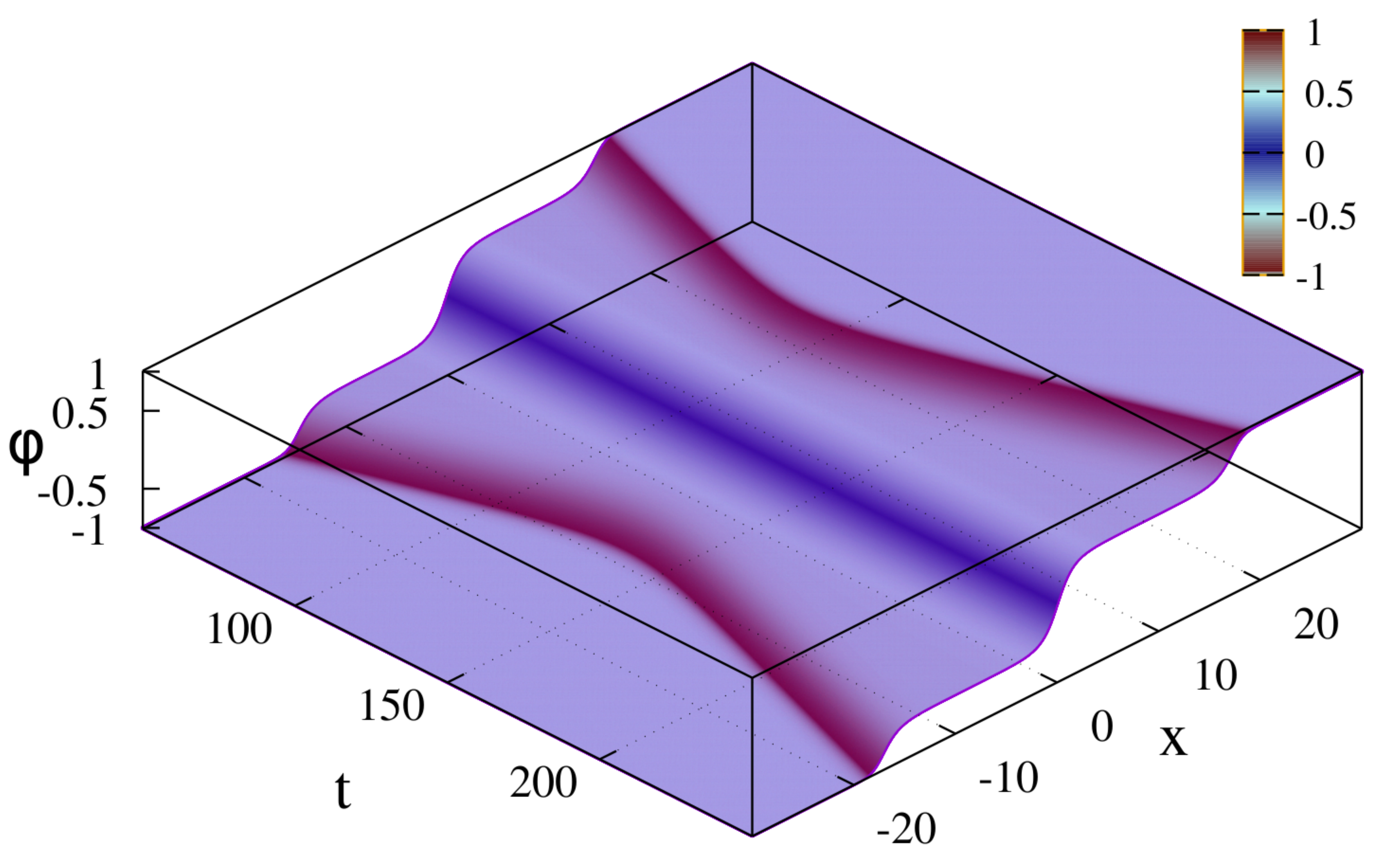}\label{fig:V01V00Vm01Xm200X00Xp200Fieldm10m05p05p10collision}}
\\
  \caption{Space-time plots of three-kink collisions.}
  \label{fig:3kinkscollision}
\end{center}
\end{figure}

It is noteworthy that the initial configuration in Fig.~\ref{fig:m0505m0505collision} has odd symmetry (is antisymmetric), while further evolution does not respect this symmetry. At the same time, the antisymmetry should be preserved by the evolution. Detailed analysis shows that the final state can change significantly depending on the parameters of the numerical scheme. (For other processes studied in this work, such changes are not observed.) Apparently, Fig.~\ref{fig:m0505m0505collision} presents a process that is largely a consequence of an instability due to slight symmetry breaking in the initial data or in the numerical calculations. Thus, we can say that such a process is impossible in an exact field-theoretic model, nevertheless, something like this can happen in a real physical system which is subject to various imperfections and fluctuations. On the other hand, it is possible to break the antisymmetry of the initial configuration by hand, e.g., by slightly changing the initial position of the left kink. Such a change should also lead to a significant change in the final state. This looks quite natural, since the processes under consideration are resonant and their details are highly dependent on fine tuning.

We also performed numerical simulation of collisions in the case of initial conditions with slightly broken antisymmetry. In this case, as expected, the dependence of the final state on the parameters of the numerical scheme disappears (since the changing parameter of the initial configuration takes over the role of the symmetry breaker). Figures \ref{fig:LKM1999} and \ref{fig:LKM2001} show collisions with initial conditions that differ very little from the case of Fig.~\ref{fig:m0505m0505collision}. Figure \ref{fig:LKM1999} shows the collision for $X_1^{}=-19.99$, and Fig.~\ref{fig:LKM2001} shows the collision for $X_1^{}=-20.01$. It is interesting that, by varying the position of the left kink, we can get formation of a kink-antikink pair in the sector $(\frac{1}{2},1)$ in the final state, see Fig.~\ref{fig:LKM2001}.

    \item Another interesting process in the case of the initial configuration of the type $(-\frac{1}{2},\frac{1}{2},-\frac{1}{2},\frac{1}{2})$ is shown in Fig.~\ref{fig:V01Vm01V00Xm10X10X30Fieldm05p05m05p05collision}. This corresponds to the initial configuration \eqref{eq:initial_condition} with $n=3$, $s_1^{}=s_3^{}=(-\frac{1}{2},\frac{1}{2})$, $s_2^{}=(\frac{1}{2},-\frac{1}{2})$, $C_3^{}=0$, $X_1^{}=-40$, $X_2^{}=-20$, $X_3^{}=0$, $v_1^{}=-v_2^{}=0.1$, $v_3^{}=0$. First, kink and antikink initially moving towards each other, collide, thus forming a pair of escaping oscillons. After that, one of the oscillons collides at the origin with the static kink $(-\frac{1}{2},\frac{1}{2})$. As a result, the oscillon passes through the kink and keeps moving in the same direction. In this case, the vacuum, around which the field oscillates in the oscillon, changes. Because of the collision with the oscillon, the kink acquires low speed in the direction of increasing $x$, $v_{\rm 2f}^{}\approx 0.031$.

    \item Next, we considered a kink-antikink-kink collision in the case of the initial configuration of the type $(\frac{1}{2},1,\frac{1}{2},1)$. We used the initial configuration \eqref{eq:initial_condition} with $n=3$, $s_1^{}=s_3^{}=(\frac{1}{2},1)$, $s_2^{}=(1,\frac{1}{2})$, $C_3^{}=-\frac{3}{2}$, $-X_1^{}=X_3^{}=20$, $X_2^{}=-2.2926$, $v_1^{}=-v_3^{}=0.1$, $v_2^{}=0$. Initial positions of the kinks are fine-tuned in order to provide their collision in one point; hence the energy density in the collision point is maximized. The space-time picture of the collision is shown in Fig.~\ref{fig:05100510collision}. In the final state, we observe radiation produced in the kink-antikink annihilation and a fast moving kink in the direction of the $x$-axis. Interestingly, part of the energy of the annihilated kink-antikink pair is converted into the kinetic energy of the escaping kink, and its final velocity $v_{\rm 3f}^{}\approx 0.712$ is approximately seven times higher than the initial one.

    \item To simulate antikink-kink-kink collisions for the initial configuration of the type $(\frac{1}{2},-\frac{1}{2},\frac{1}{2},1)$ we used the initial condition \eqref{eq:initial_condition} with $n=3$, $s_1^{}=(\frac{1}{2},-\frac{1}{2})$, $s_2^{}=(-\frac{1}{2},\frac{1}{2})$, $s_3^{}=(\frac{1}{2},1)$, $C_3^{}=0$, $X_1^{}=-40$, $X_2^{}=-20$, $X_3^{}=0$, $v_1^{}=-v_2^{}=0.1$, $v_3^{}=0$. First, the antikink $(\frac{1}{2},-\frac{1}{2})$ and the kink $(-\frac{1}{2},\frac{1}{2})$ collide with each other, producing a pair of fast escaping oscillons. Then, one of the oscillons collides with the static kink $(\frac{1}{2},1)$, and as a result, the kink gets a high velocity $v_{\rm 3f}^{}\approx 0.927$. After the collision with the kink, the oscillon significantly reduces the amplitude of field oscillations, changes direction of its motion, and escapes with a high velocity in the direction of decreasing $x$; see Fig.~\ref{fig:V01Vm01V00Xm10X10X30Fieldp05m05p05p10collision}.

    \item We observed another new phenomenon in antikink-kink-kink collisions for the initial configuration of the type $(-\frac{1}{2},-1,-\frac{1}{2},\frac{1}{2})$. In this case we used the initial condition \eqref{eq:initial_condition} with $n=3$, $s_1^{}=(-\frac{1}{2},-1)$, $s_2^{}=(-1,-\frac{1}{2})$, $s_3^{}=(-\frac{1}{2},\frac{1}{2})$, $C_3^{}=\frac{3}{2}$, $X_1^{}=-40$, $X_2^{}=-28$, $X_3^{}=40$, $v_1^{}=0.84$, $v_2^{}=0.65$, $v_3^{}=-0.7$. With such initial data, all three solitary waves collide at the origin. As a result of the collision, the kink $(-\frac{1}{2},\frac{1}{2})$ slightly changes its speed, from $v_3^{}=-0.7$ to $v_{\rm 1f}^{}\approx -0.619$. In this case, the antikink-kink pair in the sector $(-1,-\frac{1}{2})$ annihilates, and in the final state a kink-antikink pair in the sector $(\frac{1}{2},1)$ is observed instead, see Fig.~\ref{fig:V084V065Vm07Xm400Xm280Xp400Fieldm05m10m05p05collision}.

    \item Finally, in the case of kink-kink-kink collision for the initial configuration of the type $(-1,-\frac{1}{2},\frac{1}{2},1)$ we observed elastic reflection of light kinks $(-1,-\frac{1}{2})$ and $(\frac{1}{2},1)$ from the heavy kink $(-\frac{1}{2},\frac{1}{2})$. In this case, we used the initial condition \eqref{eq:initial_condition} with $n=3$, $s_1^{}=(-1,-\frac{1}{2})$, $s_2^{}=(-\frac{1}{2},\frac{1}{2})$, $s_3^{}=(\frac{1}{2},1)$, $C_3^{}=0$, $-X_1^{}=X_3^{}=20$, $X_2^{}=0$, $v_1^{}=-v_3^{}=0.1$, $v_2^{}=0$. The initial conditions are constructed so that the center of mass of the entire system is at rest at the origin, i.e., at the center of the heavy static kink $(-\frac{1}{2},\frac{1}{2})$. As a result of repulsion between the kinks, an elastic reflection of the incident kinks occurs, see Fig.~\ref{fig:V01V00Vm01Xm200X00Xp200Fieldm10m05p05p10collision}. In this case, no radiation of energy in the form of small-amplitude waves is observed.

\end{enumerate}

The above information on collisions of three kinks/antikinks (except the auxiliary cases shown in Figs.~\ref{fig:LKM1999} and \ref{fig:LKM2001}) is summarised in Table \ref{tab:Table_2}.

\renewcommand{\arraystretch}{0.9}
\begin{table}[t!]
\begin{ruledtabular}
\caption{Brief summary of processes observed in the collisions of three kinks/antikinks.}
\label{tab:Table_2}
\begin{tabular}{ccccc}
type & ``reaction'' & initial
velocities & final velocities & figure\\
\hline
\vphantom{$\displaystyle\frac{1}{2}$} $(-\frac{1}{2},\frac{1}{2},-\frac{1}{2},\frac{1}{2})$ & $K\bar{K}K\to \bar{k}kK$ & $v_1^{}=-v_3^{}=0.1$, & $v_{\rm 1f}^{}=-0.834$, $v_{\rm 2f}^{}=-0.653$, & \ref{fig:m0505m0505collision}\\
 & & $v_2^{}=0$ & $v_{\rm 3f}^{}=0.698$ & \\
\vphantom{$\displaystyle\frac{1}{2}$} $(-\frac{1}{2},\frac{1}{2},-\frac{1}{2},\frac{1}{2})$ & $K\bar{K}K\to oKo$ & $v_1^{}=-v_2^{}=0.1$ & $v_{\rm 2f}^{}=0.031$ & \ref{fig:V01Vm01V00Xm10X10X30Fieldm05p05m05p05collision}\\
 & & $v_3^{}=0$ & & \\
\vphantom{$\displaystyle\frac{1}{2}$} $(\frac{1}{2},1,\frac{1}{2},1)$ & $k\bar{k}k\to k$ & $v_1^{}=-v_3^{}=0.1$ & $v_{\rm f}^{}=0.712$ & \ref{fig:05100510collision}\\
 & & $v_2^{}=0$ & & \\
\vphantom{$\displaystyle\frac{1}{2}$} $(\frac{1}{2},-\frac{1}{2},\frac{1}{2},1)$ & $\bar{K}Kk\to ook$ & $v_1^{}=-v_2^{}=0.1$ & $v_{\rm 3f}^{}=0.927$ & \ref{fig:V01Vm01V00Xm10X10X30Fieldp05m05p05p10collision}\\
 & & $v_3^{}=0$ & & \\
\vphantom{$\displaystyle\frac{1}{2}$} $(-\frac{1}{2},-1,-\frac{1}{2},\frac{1}{2})$ & $\bar{k}kK\to Kk\bar{k}$ & $v_1^{}=0.84$, $v_2^{}=0.65$ & $v_{\rm 1f}^{}=-0.619$, $v_{\rm 2f}^{}=0.288$ & \ref{fig:V084V065Vm07Xm400Xm280Xp400Fieldm05m10m05p05collision}\\
 & & $v_3^{}=-0.7$ & $v_{\rm 3f}^{}=0.844$ & \\
\vphantom{$\displaystyle\frac{1}{2}$} $(-1,-\frac{1}{2},\frac{1}{2},1)$ & $kKk\to kKk$ & $v_1^{}=-v_3^{}=0.1$ & $v_{\rm 1f}^{}=-v_{\rm 3f}^{}=-0.1$ & \ref{fig:V01V00Vm01Xm200X00Xp200Fieldm10m05p05p10collision}\\
 & & $v_2^{}=0$ & $v_{\rm 2f}^{}=0$ & \\
\end{tabular}
\end{ruledtabular}
\end{table}
\renewcommand{\arraystretch}{1}

\subsection{Four kinks}
\label{sec:Results.Four}

We have performed numerical simulations of the following four-kink processes.
\begin{enumerate}

    \item Collision of two kinks and two antikinks belonging to the topological sector $(-\frac{1}{2},\frac{1}{2})$. In this case, we used the initial condition \eqref{eq:initial_condition} with $n=4$, $s_1^{}=s_3^{}=(-\frac{1}{2},\frac{1}{2})$, $s_2^{}=s_4^{}=(\frac{1}{2},-\frac{1}{2})$, $C_4^{}=-\frac{1}{2}$, $-X_1^{}=X_4^{}=24.60619$, $-X_2^{}=X_3^{}=10$, $v_1^{}=-v_4^{}=0.1$, $v_2^{}=-v_3^{}=0.05$. As in previous simulations, initial positions and initial velocities of kinks are chosen so that all solitons arrive at the collision point simultaneously. As one can see from  Fig.~\ref{fig:m0505m0505m05collision}, the result of the collision is two kink-antikink pairs belonging to the topological sector $(-1,-\frac{1}{2})$ that are moving away from each other.
\begin{figure}[t!]
\begin{center}
  \centering
      \subfigure[\:$(-\frac{1}{2},\frac{1}{2},-\frac{1}{2},\frac{1}{2},-\frac{1}{2})$]{\includegraphics[width=0.32
 \textwidth]
{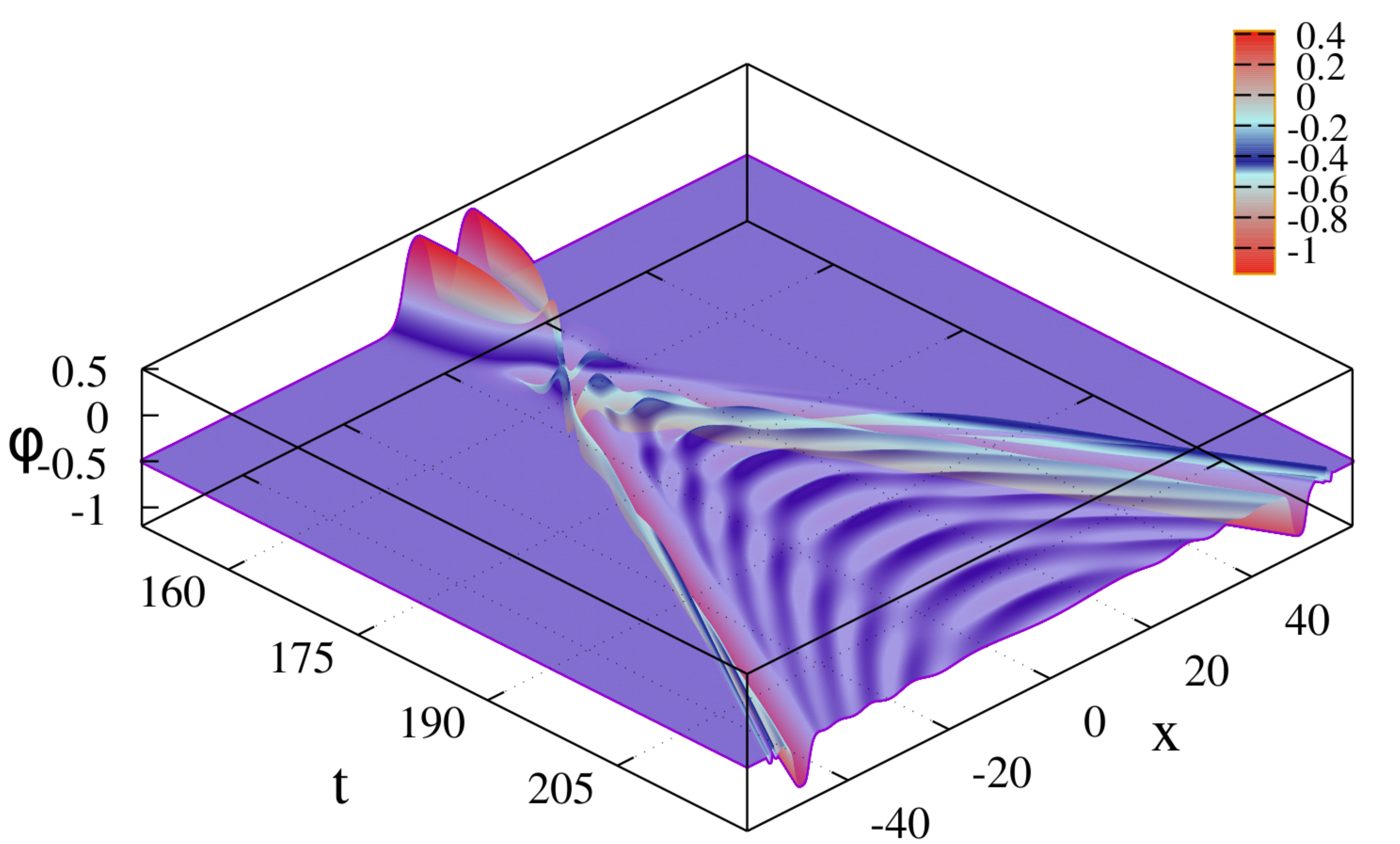}\label{fig:m0505m0505m05collision}}
      \subfigure[\:$(\frac{1}{2},1,\frac{1}{2},1,\frac{1}{2})$]{\includegraphics[width=0.32
 \textwidth]
{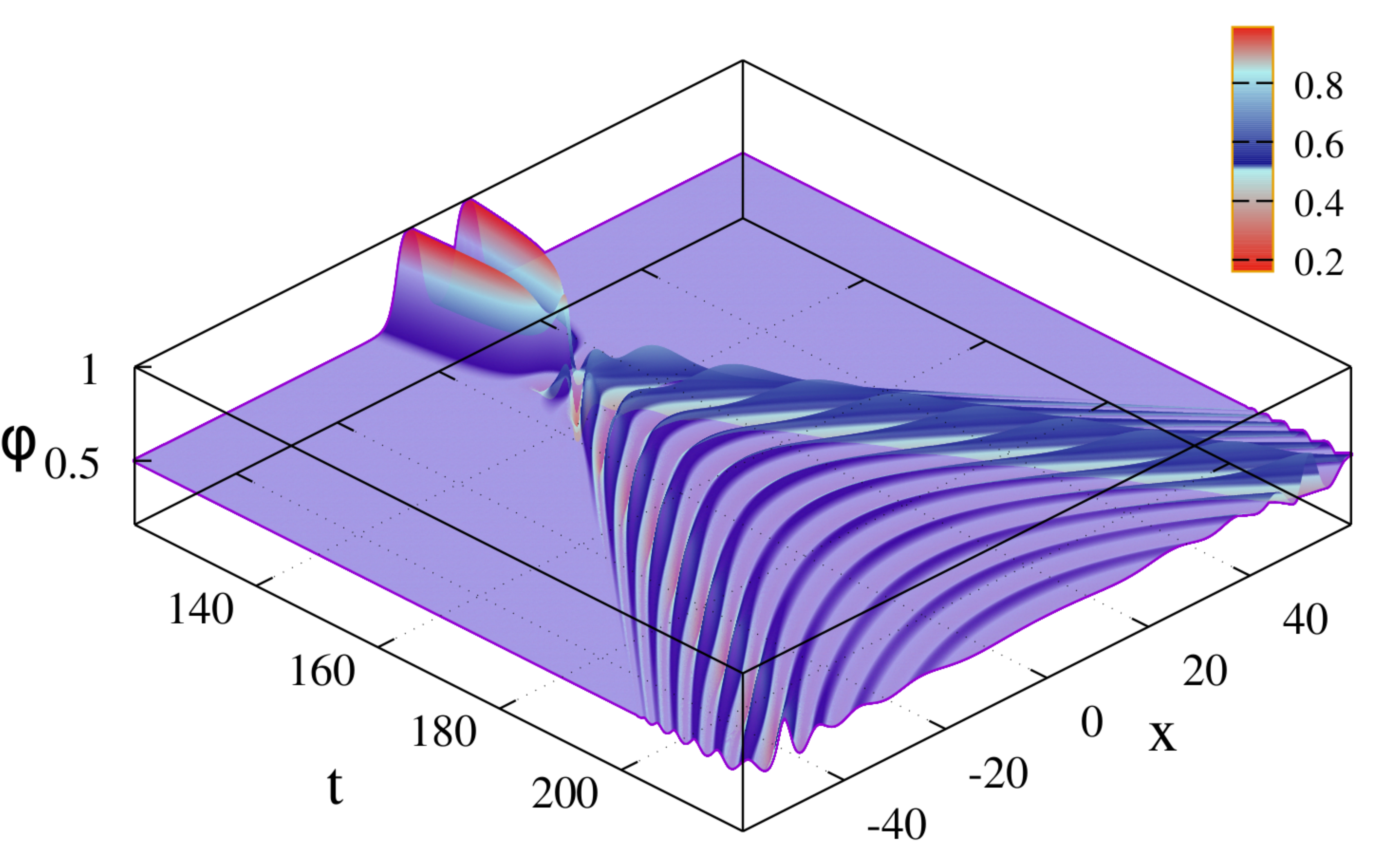}\label{fig:0510051005collision}}
    \subfigure[\:$(1,\frac{1}{2},1,\frac{1}{2},1)$]{\includegraphics[width=0.32
 \textwidth]
{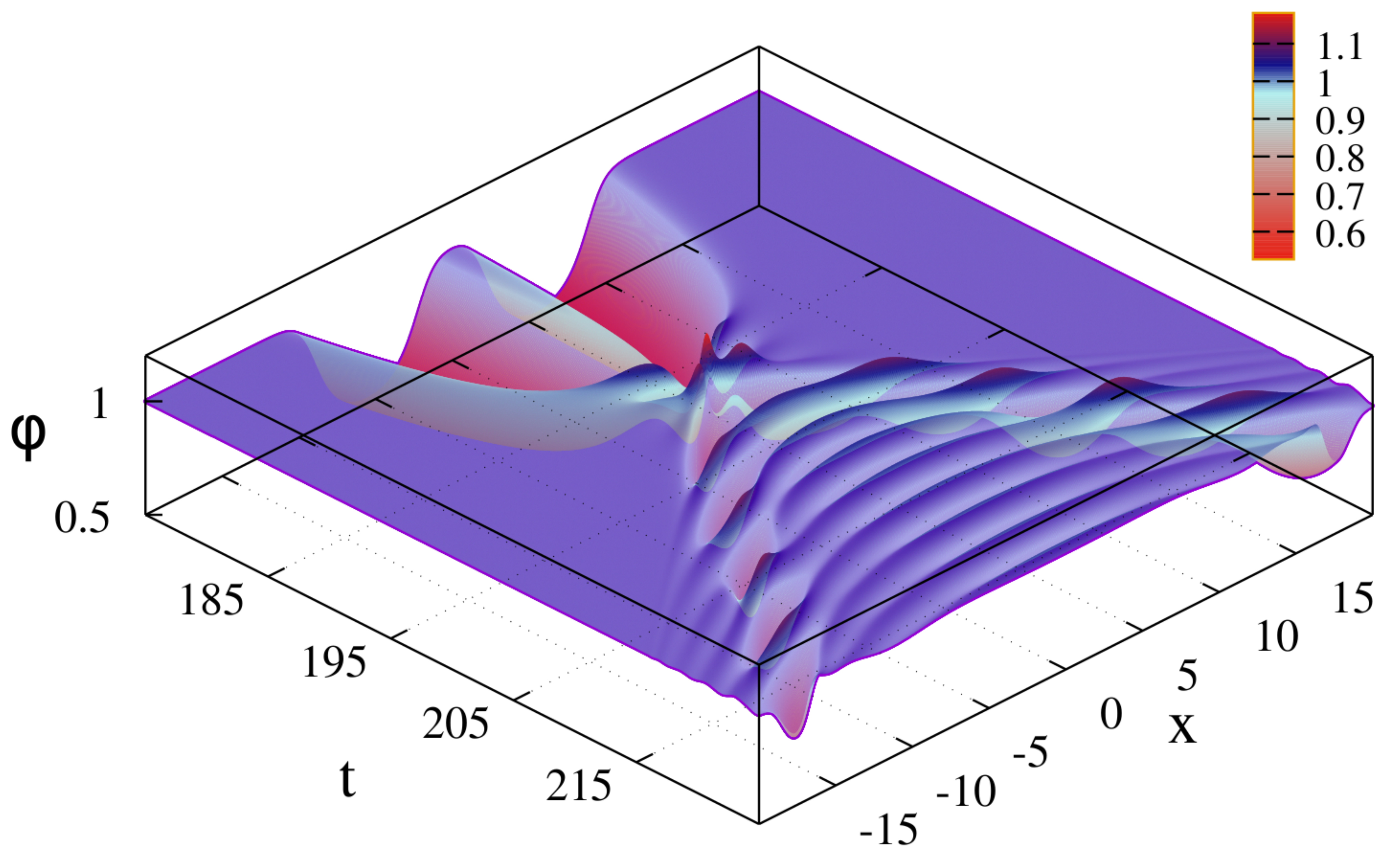}\label{fig:1005100510collision}}
  \\
  \caption{Space-time plots of four-kink collisions.}
  \label{fig:4kinkscolision}
\end{center}
\end{figure}
    It is interesting that the relative velocities of the kink and antikink in each pair are very small, they are equal to about 0.007. With a slight change in the initial conditions, the relative velocities of the kink and antikink in each pair increase. For example, for $-X_1^{}=X_4^{}=24.606215$ (and all other parameters are unchanged), the relative velocities of the kink and antikink are approximately 0.07.

    \item Collision of two kinks and two antikinks belonging to the topological sector $(\frac{1}{2},1)$. In this case, we used the initial condition \eqref{eq:initial_condition} with $n=4$, $s_1^{}=s_3^{}=(\frac{1}{2},1)$, $s_2^{}=s_4^{}=(1,\frac{1}{2})$, $C_4^{}=-\frac{5}{2}$, $-X_1^{}=X_4^{}=19.91705$, $-X_2^{}=X_3^{}=10$, $v_1^{}=-v_4^{}=0.1$, $v_2^{}=-v_3^{}=0.05$. As usual, the initial data is selected in such a way that the collision of all four waves occurs at the same point. The space-time picture of the collision is shown in Fig.~\ref{fig:0510051005collision}. It is seen that as a result of the collision, energy is generated, which is emitted in the form of small-amplitude waves.

    \item Finally, we performed a numerical simulation of the collision of the same two kinks and two antikinks belonging to the topological sector $(\frac{1}{2},1)$, but arranged on the axis in a different order. In this case, we used the initial condition \eqref{eq:initial_condition} with $n=4$, $s_1^{}=s_3^{}=(1,\frac{1}{2})$, $s_2^{}=s_4^{}=(\frac{1}{2},1)$, $C_4^{}=-2$, $-X_1^{}=X_4^{}=27.654600009$, $-X_2^{}=X_3^{}=10$, $v_1^{}=-v_4^{}=0.1$, $v_2^{}=-v_3^{}=0.05$. All four waves collided at one point. The space-time picture of the collision is shown in Fig.~\ref{fig:1005100510collision}. It can be seen that as a result of the collision, two escaping oscillons are formed.

\end{enumerate}

A brief summary of the above four-kink collisions is presented in Table \ref{tab:Table_3}.

\begin{table}[t!]
\begin{ruledtabular}
\caption{Brief summary of processes observed in the collisions of four kinks/antikinks.}
\label{tab:Table_3}
\begin{tabular}{ccccc}
type & ``reaction'' & initial
velocities & final velocities & figure\\
\hline
\vphantom{$\displaystyle\frac{1}{2}$} $(-\frac{1}{2},\frac{1}{2},-\frac{1}{2},\frac{1}{2},-\frac{1}{2})$ & $K\bar{K}K\bar{K}\to \bar{k}k\bar{k}k$ & $v_1^{}=-v_4^{}=0.1$, & $-v_{\rm 1f}^{}=v_{\rm 4f}^{}=0.876$, & \ref{fig:m0505m0505m05collision}\\
 & & $v_2^{}=-v_3^{}=0.05$ & $-v_{\rm 2f}^{}=v_{\rm 3f}^{}=0.869$ & \\
\vphantom{$\displaystyle\frac{1}{2}$} $(\frac{1}{2},1,\frac{1}{2},1,\frac{1}{2})$ & $k\bar{k}k\bar{k}\to radiation$ & $v_1^{}=-v_4^{}=0.1$, & --- & \ref{fig:0510051005collision}\\
 & & $v_2^{}=-v_3^{}=0.05$ & & \\
\vphantom{$\displaystyle\frac{1}{2}$} $(1,\frac{1}{2},1,\frac{1}{2},1)$ & $\bar{k}k\bar{k}k\to oo$ & $v_1^{}=-v_4^{}=0.1$, & --- & \ref{fig:1005100510collision}\\
 & & $v_2^{}=-v_3^{}=0.05$ & & \\

\end{tabular}
\end{ruledtabular}
\end{table}

\subsection{Collisions of oscillons}
\label{sec:Results.Oscillons}

The scattering of oscillons is of particular interest. In such events, we observed nontrivial phenomena --- the production of one or two kink-antikink pairs.

\begin{figure}[t!]
\begin{center}
  \centering
  \subfigure[]{\includegraphics[width=0.40
 \textwidth]
{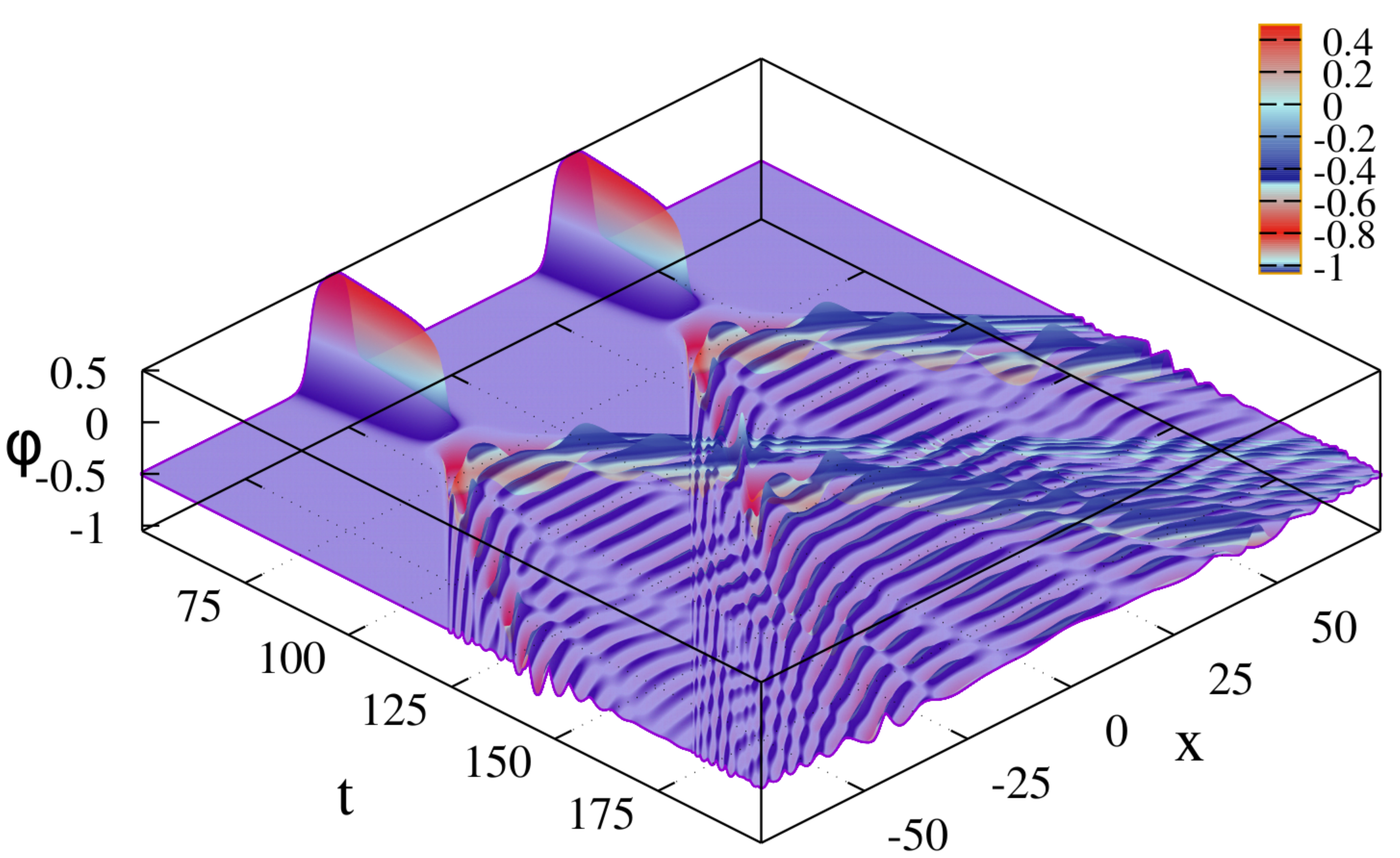}\label{fig:m0505m05F2KV01X19203920}}
  \subfigure[]{\includegraphics[width=0.40
 \textwidth]
{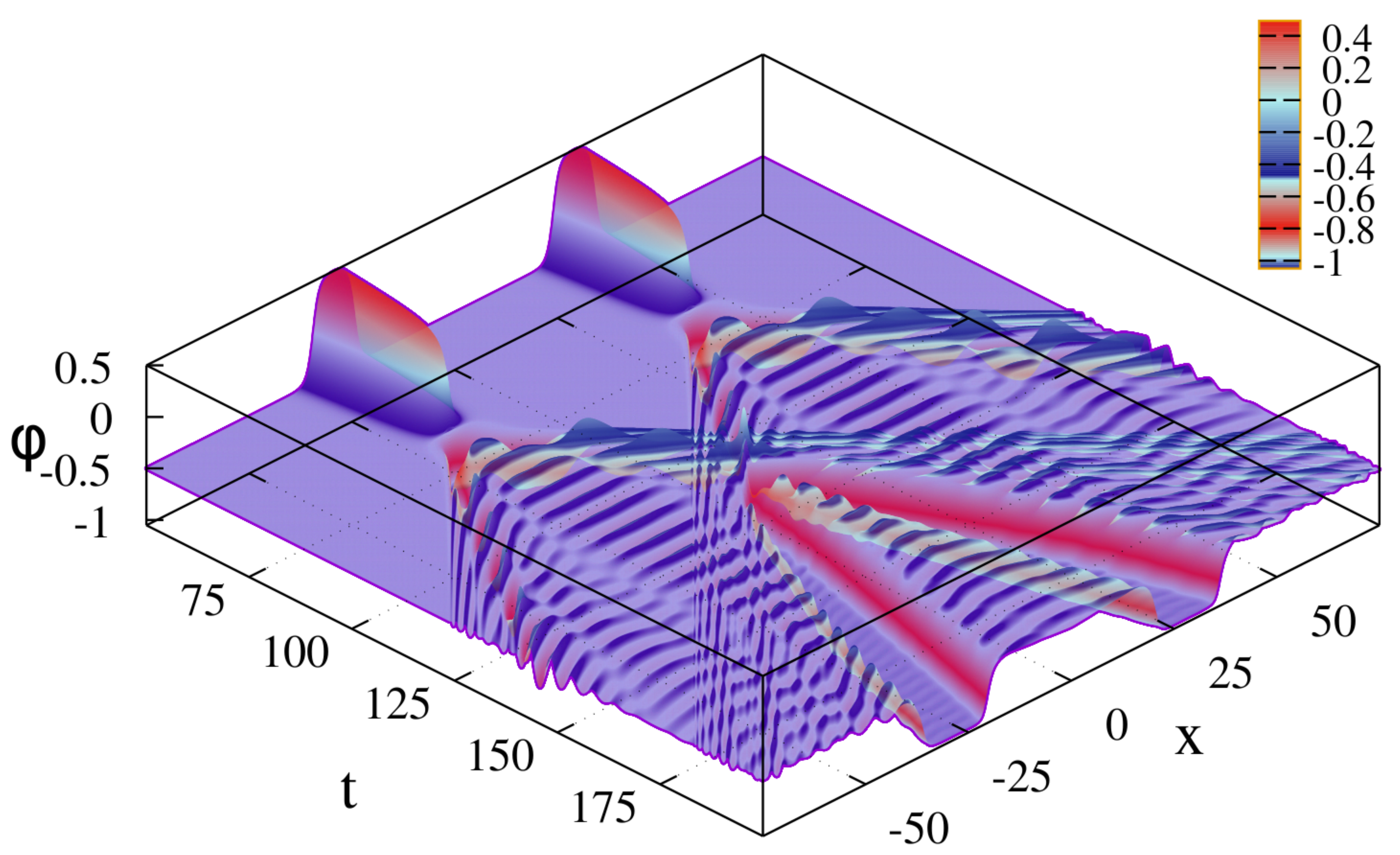}\label{fig:m0505m05F2KV01X19243924}}
\\
  \subfigure[]{\includegraphics[width=0.40
 \textwidth]
{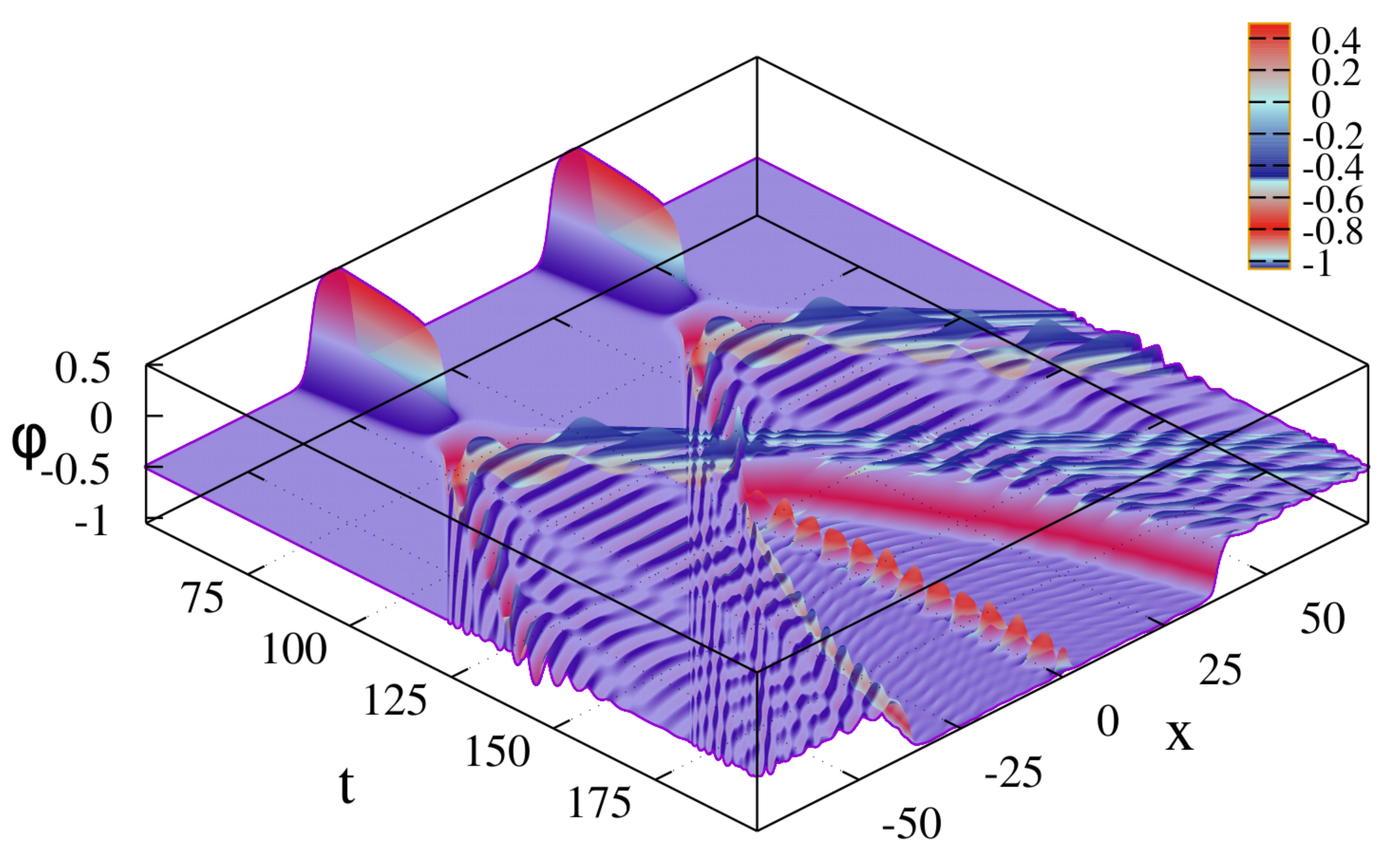}\label{fig:m0505m05F2KV01X19283928}}
    \subfigure[]{\includegraphics[width=0.40
 \textwidth]
{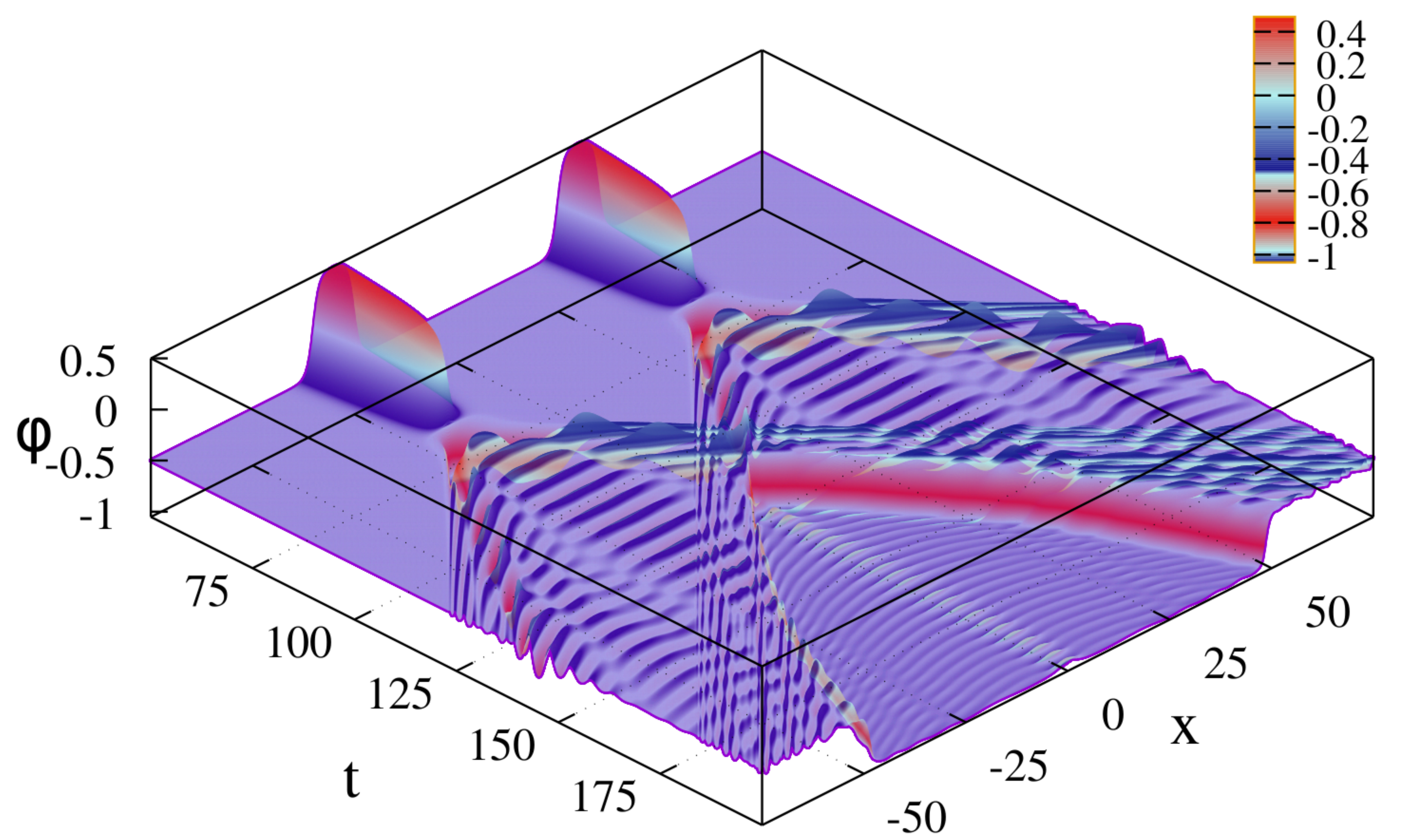}\label{fig:m0505m05F2KV01X20004000}}
  \\
  \caption{Space-time plots of oscillon-oscillon collisions.}
  \label{fig:oscillonoscillonCollision}
\end{center}
\end{figure}

At the first (preparatory) stage, we used the initial condition \eqref{eq:initial_condition} with $n=4$, $s_1^{}=s_3^{}=(-\frac{1}{2},\frac{1}{2})$, $s_2^{}=s_4^{}=(\frac{1}{2},-\frac{1}{2})$, $C_4^{}=-\frac{1}{2}$. The initial positions and initial velocities of the colliding kinks were fine-tuned in such a way as to ensure pairwise collisions of the kink and antikink. As a result of each such collision, two escaping oscillons were produced, one of which later collided with an oscillon produced from the other kink-antikink pair, see Fig.~\ref{fig:oscillonoscillonCollision}.

In all cases, we used the same set of initial kink velocities: $v_1^{}=-v_2^{}=v_3^{}=-v_4^{}=0.1$. By changing the initial positions of the colliding kinks, one can achieve different relative phases of the internal field oscillations in the colliding oscillons. This, in turn, leads to different final states in collisions of the oscillons.

First of all, at $-X_1^{}=X_4^{}=39.20$, $-X_2^{}=X_3^{}=19.20$, as a result of the collision of oscillons, we observed radiation propagating in the form of small-amplitude waves, as well as moving oscillating lumps of energy, similar to the original oscillons, but with a smaller amplitude of oscillations, see Fig.~\ref{fig:m0505m05F2KV01X19203920}.

Next, at $-X_1^{}=X_4^{}=39.24$, $-X_2^{}=X_3^{}=19.24$, as a result of the collision of oscillons, two antikink-kink pairs are formed in the topological sector $(-1,-\frac{1}{2})$, see Fig.~\ref{fig:m0505m05F2KV01X19243924}. The relative velocities of the antikink and kink in each pair are small. The centers of mass of the pairs move away from the collision point of oscillons (i.e., from the origin).

Further, in Figs.~\ref{fig:m0505m05F2KV01X19283928} and \ref{fig:m0505m05F2KV01X20004000} the scenarios for the initial positions $-X_1^{}=X_4^{}=39.28$, $-X_2^{}=X_3^{}=19.28$ and $-X_1^{}=X_4^{}=40.00$, $-X_2^{}=X_3^{}=20.00$, respectively, are shown. In both cases, as a result of the collision of oscillons, one antikink-kink pair is produced in the sector $(-1,-\frac{1}{2})$, and also a static oscillating structure is formed at the origin, which can be classified as some kind of oscillon. Comparing scenarios of Figs.~\ref{fig:m0505m05F2KV01X19283928} and \ref{fig:m0505m05F2KV01X20004000}, we conclude that the oscillon takes more energy in the first case than in the second. As the natural consequence, in the first case, the velocities of the escaping antikink and kink turn out to be somewhat lower than in the second case.

Additional numerical calculations show that the described processes depend rather strongly on the initial conditions: even with a very small change in the initial data, the final state can change quite significantly. This is a completely natural situation, taking into account the resonant nature of the processes.

\section{Deeper study of two-kink collisions}
\label{sec:K-AK}

First, we have performed a detailed numerical analysis of the kink-antikink collisions in the topological sector $(-\frac{1}{2},\frac{1}{2})$. The critical velocity is apparently very high, i.e., is close to one. We performed numerical simulation of kink-antikink collisions up to the initial velocity 0.94 and observed only kink annihilation with the formation of oscillons. We did not find any escape windows (an escape window is the range of initial velocities, within which the kinks escape to spatial infinities after two or more collisions). This is consistent with the fact that there are no vibrational modes in the excitation spectrum of the kink and, in addition, the kink's stability potential is symmetric (for more details of the kink's stability analysis see, e.g., \cite{Gani.JHEP.2015}).

Second, we have studied the kink-antikink collisions in the topological sector $(\frac{1}{2},1)$. We have found the critical velocity $v_{\rm cr}\approx 0.88$: at initial velocities $v_{\rm in}>v_{\rm cr}$ kink and antikink escape to spatial infinities after one collision; at $v_{\rm in}<v_{\rm cr}$ they capture each other and form a long-lived bound state, which then decays slowly radiating small-amplitude waves. As in the kink-antikink collisions in the sector $(-\frac{1}{2},\frac{1}{2})$, we did not observe any escape windows. This is consistent with the fact that the discrete spectrum of the kink stability potential contains only zero mode. Moreover, although in this case the stability potential of the kink is asymmetric, the stability potential of the kink-antikink system as a whole does not form a potential well.

\begin{figure}[t!]
\begin{center}
  \centering
    {\includegraphics[width=0.5
 \textwidth]
{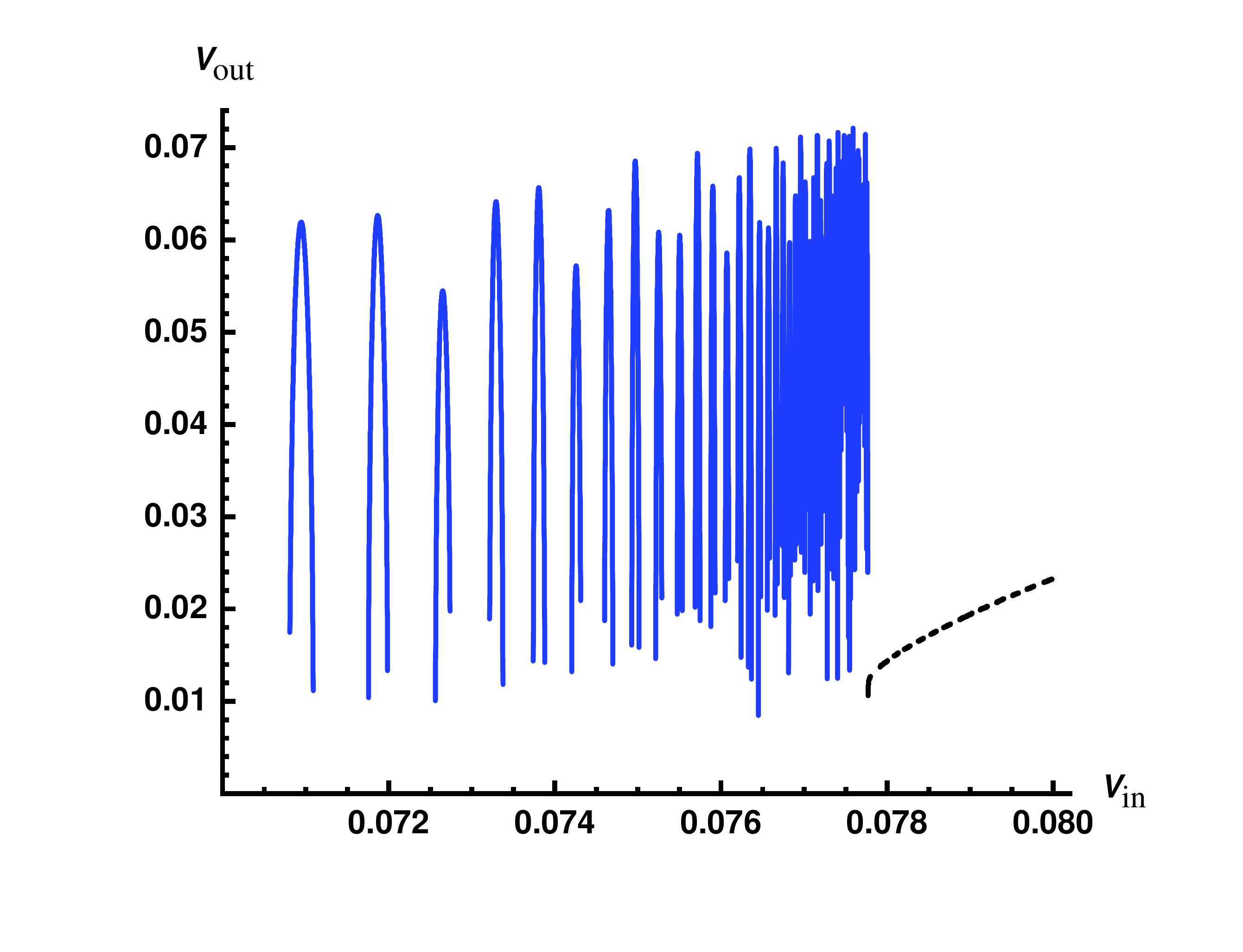}\label{Figs/onetwobounces}}
  \caption{Plot of the escape velocity $v_{\rm out}^{}$ as a function of the initial velocity $v_{\rm in}^{}$. The two-bounce escape windows (blue solid curves) and one-bounce window (black dashed curve) are shown. The antikink and kink initial positions are $X_1^{}=-X_2^{}=-10$; the initial velocity step is $\delta v_{\rm in}^{}=0.000001$.}
  \label{fig:Bounces}
\end{center}
\end{figure}

\begin{figure}[t!]
\begin{center}
  \centering
  \subfigure[\:$v_{\rm in}^{}=0.075000$ (two-bounce window)]{\includegraphics[width=0.40
 \textwidth]
{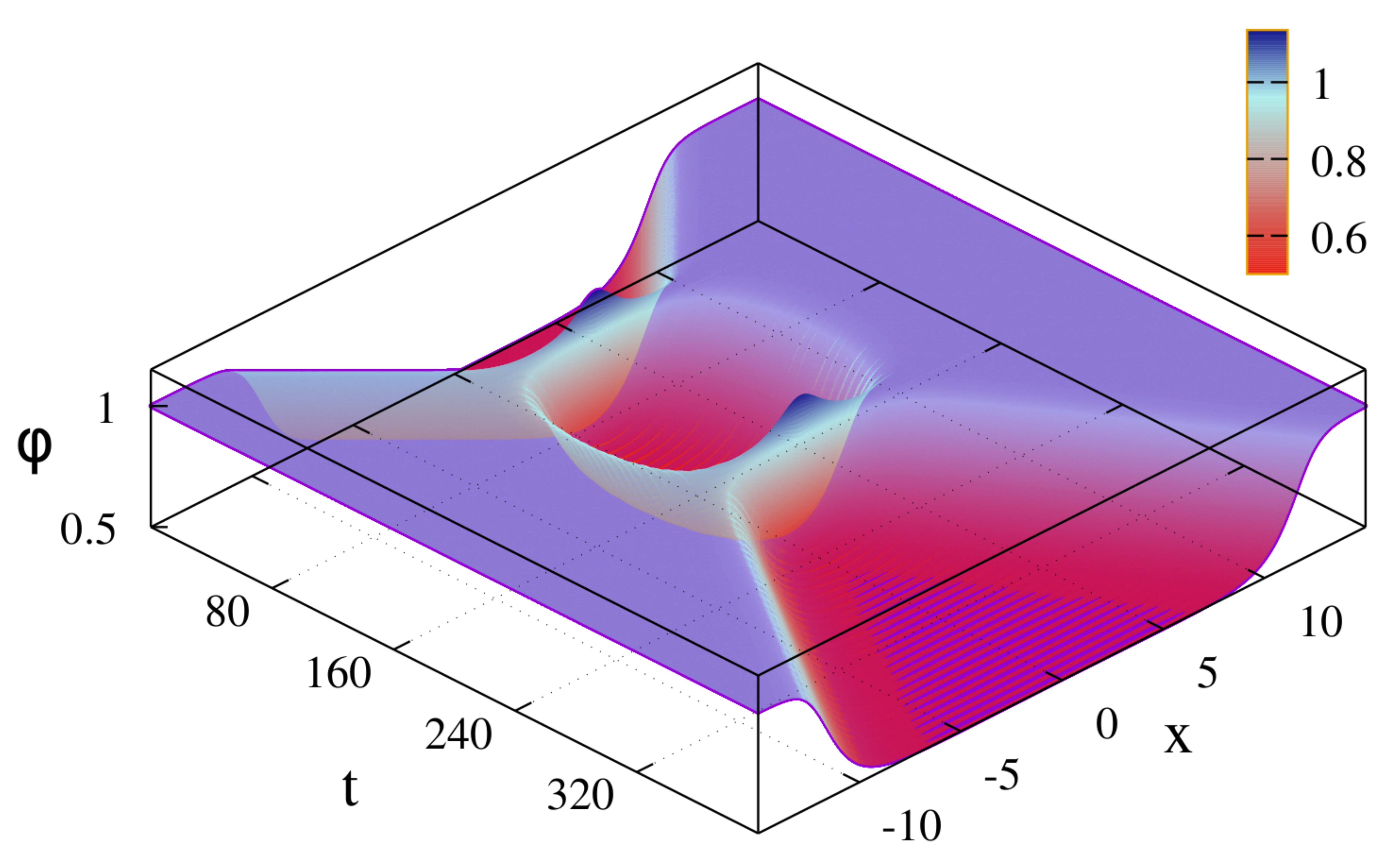}\label{fig:100510F2KV00750X10}}
  \subfigure[\:$v_{\rm in}^{}=0.075300$ (three-bounce window)]{\includegraphics[width=0.40
 \textwidth]
{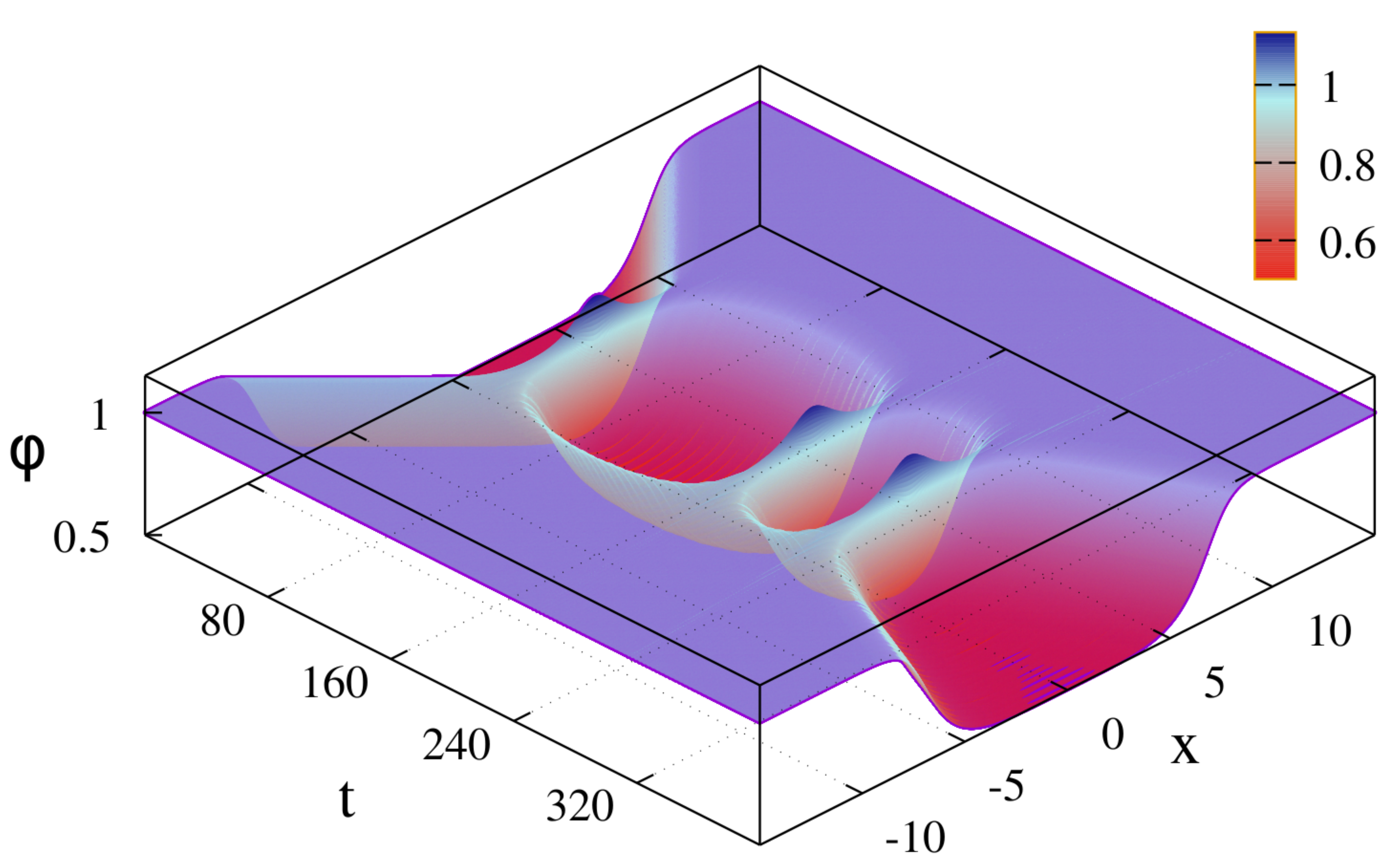}\label{fig:100510F2KV00753X10}}
\\
  \subfigure[\:$v_{\rm in}^{}=0.070277$ (four-bounce window)]{\includegraphics[width=0.40
 \textwidth]
{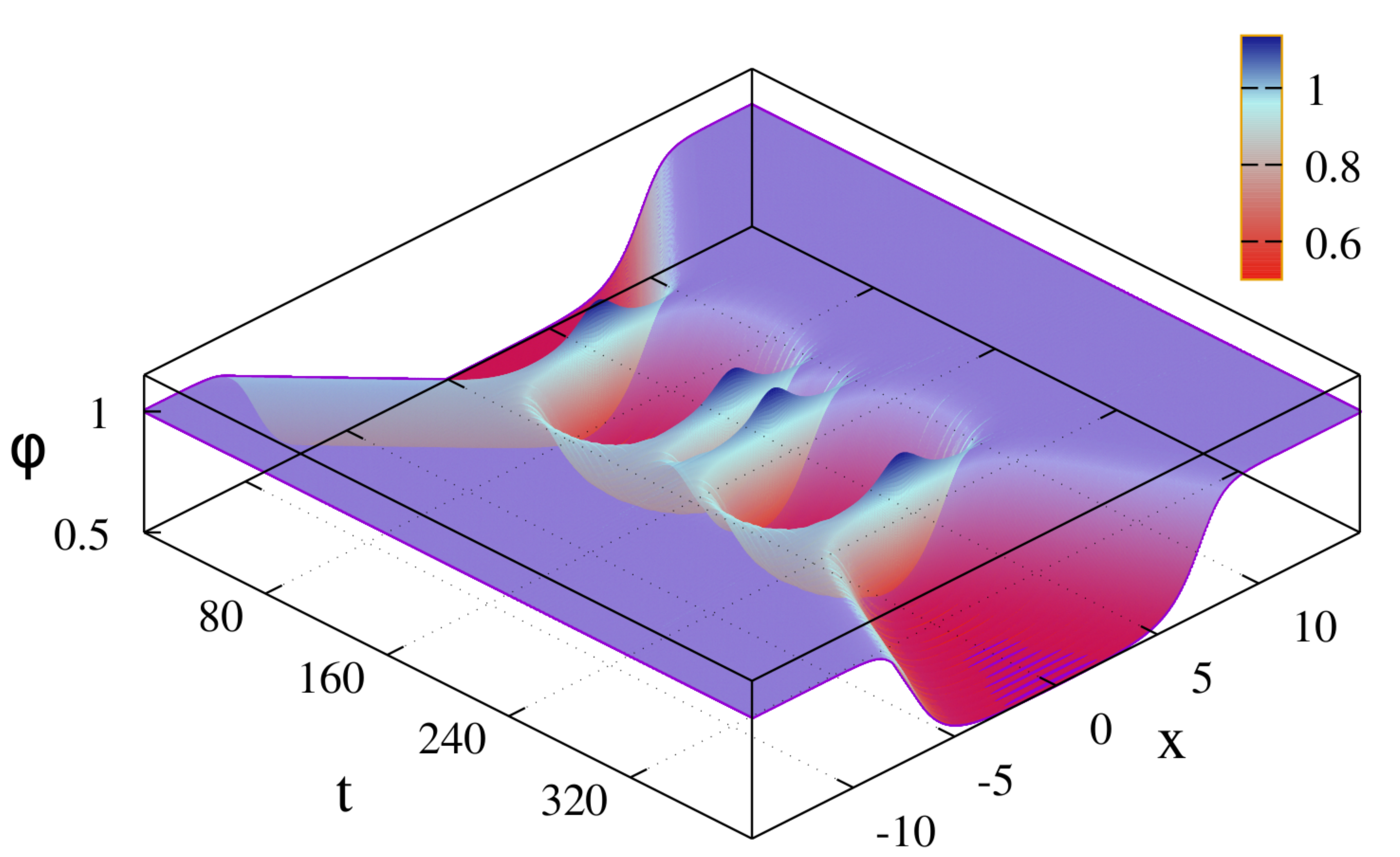}\label{fig:100510F2KV0070277X10}}
  \subfigure[\:$v_{\rm in}^{}=0.070240$ (five-bounce window)]{\includegraphics[width=0.40
 \textwidth]
{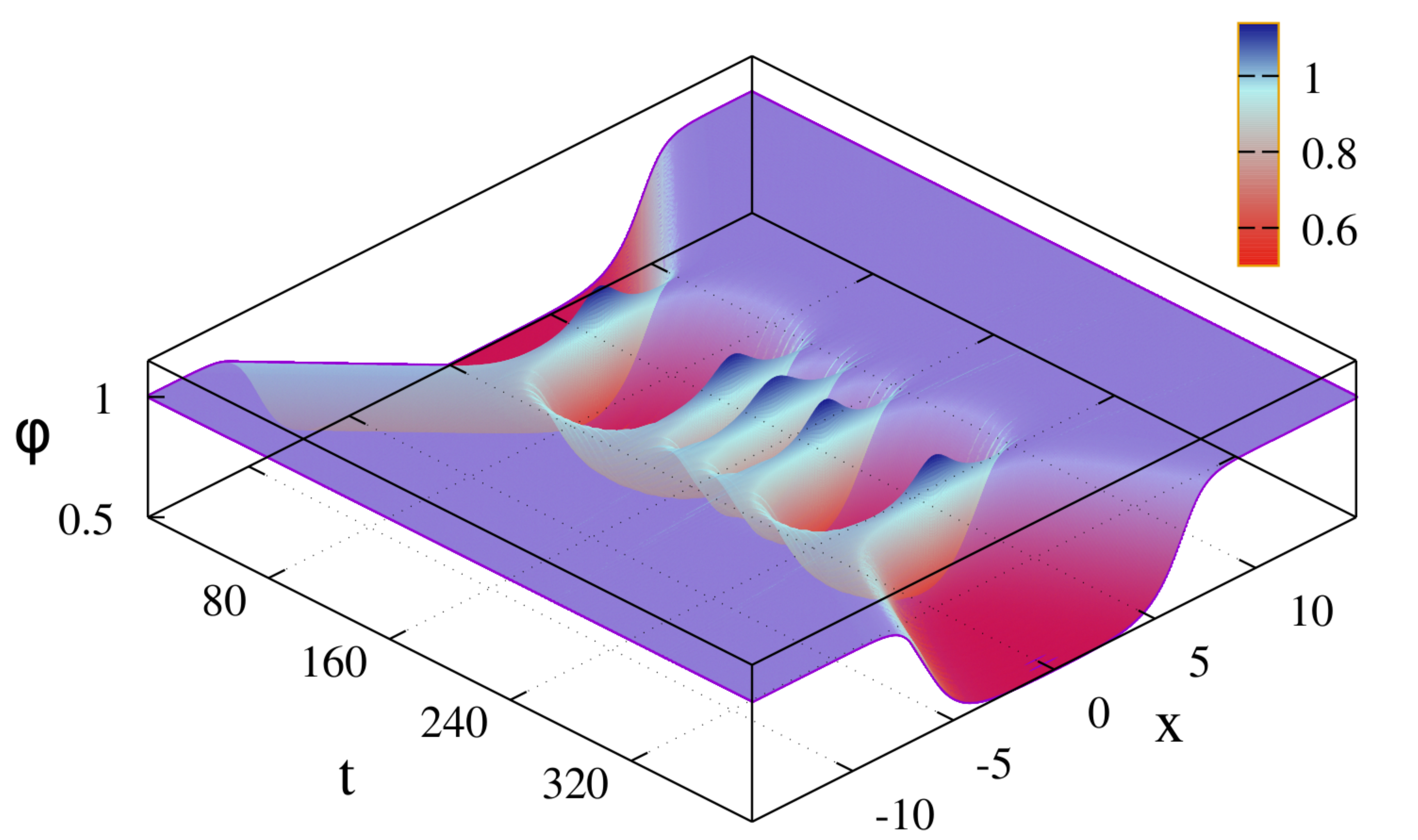}\label{fig:100510F2KV0070240X10}}
\\
  \caption{Space-time plots of the antikink-kink collisions in the sector $(\frac{1}{2},1)$ for different initial velocities $v_{\rm in}^{}$ belonging to the escape windows. The initial positions of the kinks in all cases are $X_1^{}=-X_2^{}=10$.}
  \label{fig:2kdv100510}
\end{center}
\end{figure}

Third, we have performed numerical simulation of the antikink-kink collisions in the same sector $(\frac{1}{2},1)$. This case is the most interesting of all discussed in this section. Despite the absence of a vibrational mode in the kink excitation spectrum, in antikink-kink scattering we found a rich structure of escape windows. Of course, all the escape windows are located below the critical velocity, which in this case we found to be $v_{\rm cr}\approx 0.07777$. The structure of the two-bounce windows is shown in Fig.~\ref{fig:Bounces}.
It is seen that when approaching the critical velocity, the two-bounce windows become narrower. We also found many three-bounce, four-bounce and so on escape windows (some selected examples of the field dynamics within escape windows are presented in Fig.~\ref{fig:2kdv100510}). However, their detailed study is a grueling procedure and is beyond the scope of our work.

\section{Discussion and Conclusion}
\label{sec:Conclusion}

We have carried out a numerical study of exotic processes in collisions of several kinks of the $\varphi^8$ model. Let us briefly summarise our main findings.

{\it Change of the topological sector.} This phenomenon looks rather trivial in a model with a periodic potential, such as, e.g., the double sine-Gordon \cite{Gani.EPJC.2018,Gani.EPJC.2019}. In the case of the periodic potential, the model has an infinite number of topological sectors. The kinks in these sectors do not differ in anything, except for the addition of a constant that is a multiple of the potential's period. In this case, the transition of a kink from one topological sector to another can occur due to kinks passage through each other. In the model we have considered, the situation is fundamentally different --- there are three neighboring topological sectors in which the kink solutions differ. As a result, the transition of a kink or a kink-antikink pair from one sector to another is a nontrivial process. We observed the transition of a kink-antikink pair from one topological sector to another in the collision of three kinks in cases of initial configurations of the types $(-\frac{1}{2},\frac{1}{2},-\frac{1}{2},\frac{1}{2})$ and $(-\frac{1}{2},-1,-\frac{1}{2},\frac{1}{2})$, Figs.~\ref{fig:m0505m0505collision} and \ref{fig:V084V065Vm07Xm400Xm280Xp400Fieldm05m10m05p05collision}, and also in the collision of four kinks forming an initial configuration of the type $(-\frac{1}{2},\frac{1}{2},-\frac{1}{2},\frac{1}{2},-\frac{1}{2})$, Fig.~\ref{fig:m0505m0505m05collision}. As mentioned above, the process shown in Fig.~\ref{fig:m0505m0505collision} should be interpreted with great care, since it does not respect the antisymmetry of the initial configuration, see Section \ref{sec:Results.Three}.

{\it Scattering of an oscillon by a kink.} The process was observed in cases of initial configurations of the types $(-\frac{1}{2},\frac{1}{2},-\frac{1}{2},\frac{1}{2})$ and $(\frac{1}{2},-\frac{1}{2},\frac{1}{2},1)$, Figs.~\ref{fig:V01Vm01V00Xm10X10X30Fieldm05p05m05p05collision} and \ref{fig:V01Vm01V00Xm10X10X30Fieldp05m05p05p10collision}. In the first case, the oscillon about the vacuum $-\frac{1}{2}$ passes through the kink $(-\frac{1}{2},\frac{1}{2})$ almost unhindered and falls into the vacuum $+\frac{1}{2}$. In this case, the kink almost does not change its state. In the second case, the collision of the oscillon about the vacuum $\frac{1}{2}$ and the kink $(\frac{1}{2},1)$ results in a bounce of the oscillon and a significant momentum transfer to the kink.

{\it Production of kink-antikink pairs.} Oscillons can carry energy sufficient to produce one or even two kink-antikink pairs. We have observed the production of kink-antikink pairs in collisions of two oscillons about the vacuum $-\frac{1}{2}$. In this case, depending on the relative phase of the field oscillations in the oscillons, we observed the passage of oscillons through each other, Fig.~\ref{fig:m0505m05F2KV01X19203920}, the production of two kink-antikink pairs, Fig.~\ref{fig:m0505m05F2KV01X19243924}, or the production of a kink-antikink pair and an oscillon, Figs.~\ref{fig:m0505m05F2KV01X19283928} and \ref{fig:m0505m05F2KV01X20004000}.

We have also performed a detailed numerical analysis of the kink-antikink and antikink-kink collisions. In the antikink-kink collisions in the sector $(\frac{1}{2},1)$ we have found a rich structure of escape windows.

The construction of theoretical models could become the next step in studying the phenomena that we observed experimentally. In particular, the formation of kink-antikink pairs in collisions of oscillons, apparently, strongly depends on the phases of the field oscillations in the colliding oscillons. A more detailed study on the scattering of oscillons and the construction of a theoretical model could become the subject of a separate study. In addition, the scattering of an oscillon by a kink, apparently, also strongly depends on which vacua the kink connects and on the properties of oscillons about these vacua.

\newpage

\section*{Acknowledgments}

V.A.G.\ acknowledges the support of the Russian Foundation for Basic Research under Grant No.\ 19-02-00971. The work of the MEPhI group was also supported by MEPhI within the Program ``Priority-2030''. For A.M.M., this work is supported by Islamic Azad University Quchan branch under the grant. K.J.\ thanks the Ferdowsi University of Mashhad for supporting the work by the grant No.~2/52484.


\end{document}